\documentclass[twoside,twocolumn,9pt]{article}
\usepackage{extsizes}
\usepackage[super,sort&compress,comma]{natbib} 
\usepackage[version=3]{mhchem}
\usepackage[left=1.5cm, right=1.5cm, top=1.785cm, bottom=2.0cm]{geometry}
\usepackage{balance}
\usepackage{mathptmx}
\usepackage{sectsty}
\usepackage{graphicx} 
\usepackage{lastpage}
\usepackage[format=plain,justification=justified,singlelinecheck=false,font={stretch=1.125,small,sf},labelfont=bf,labelsep=space]{caption}
\usepackage{float}
\usepackage{fancyhdr}
\usepackage{fnpos}
\usepackage[english]{babel}
\usepackage{array}
\usepackage{droidsans}
\usepackage{charter}
\usepackage[T1]{fontenc}
\usepackage[usenames,dvipsnames]{xcolor}
\usepackage{setspace}
\usepackage[compact]{titlesec}
\usepackage{hyperref}

\usepackage{epstopdf}

\definecolor{cream}{RGB}{222,217,201}

\usepackage{xcolor}
\usepackage{xspace}

\newcommand{\Cth}{${}^{13}\mathrm{C}$\xspace}
\newcommand{\Nlab}{${}^{15}\mathrm{N}$\xspace}
\newcommand{\Hone}{${}^{1}\mathrm{H}$\xspace}
\newcommand{\degC}{$^\circ$C\xspace}

\newcommand{\IrCat}{[IrCl(IMes)(COD)]\xspace}
\newcommand{\IMes}{1,3-bis(2,4,6-trimehylphenyl)imidazol-2-ylidene)\xspace}
\newcommand{\degr}{$^\circ$\xspace}
\newcommand{\note}[1]{{\color{cyan}{#1}}}

\begin{document}

\pagestyle{fancy}
\thispagestyle{plain}
\fancypagestyle{plain}{
\renewcommand{\headrulewidth}{0pt}
}

\makeFNbottom
\makeatletter
\renewcommand\LARGE{\@setfontsize\LARGE{15pt}{17}}
\renewcommand\Large{\@setfontsize\Large{12pt}{14}}
\renewcommand\large{\@setfontsize\large{10pt}{12}}
\renewcommand\footnotesize{\@setfontsize\footnotesize{7pt}{10}}
\makeatother

\renewcommand{\thefootnote}{\fnsymbol{footnote}}
\renewcommand\footnoterule{\vspace*{1pt}%
\color{cream}\hrule width 3.5in height 0.4pt \color{black}\vspace*{5pt}} 
\setcounter{secnumdepth}{5}

\makeatletter 
\renewcommand\@biblabel[1]{#1}            
\renewcommand\@makefntext[1]%
{\noindent\makebox[0pt][r]{\@thefnmark\,}#1}
\makeatother 
\renewcommand{\figurename}{\small{Fig.}~}
\sectionfont{\sffamily\Large}
\subsectionfont{\normalsize}
\subsubsectionfont{\bf}
\setstretch{1.125} 
\setlength{\skip\footins}{0.8cm}
\setlength{\footnotesep}{0.25cm}
\setlength{\jot}{10pt}
\titlespacing*{\section}{0pt}{4pt}{4pt}
\titlespacing*{\subsection}{0pt}{15pt}{1pt}

\fancyfoot{}
\fancyfoot[LO,RE]{\vspace{-7.1pt}\includegraphics[height=9pt]{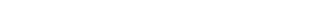}}
\fancyfoot[CO]{\vspace{-7.1pt}\hspace{11.9cm}\includegraphics{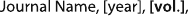}}
\fancyfoot[CE]{\vspace{-7.2pt}\hspace{-13.2cm}\includegraphics{head_foot/RF}}
\fancyfoot[RO]{\footnotesize{\sffamily{1--\pageref{LastPage} ~\textbar  \hspace{2pt}\thepage}}}
\fancyfoot[LE]{\footnotesize{\sffamily{\thepage~\textbar\hspace{4.65cm} 1--\pageref{LastPage}}}}
\fancyhead{}
\renewcommand{\headrulewidth}{0pt} 
\renewcommand{\footrulewidth}{0pt}
\setlength{\arrayrulewidth}{1pt}
\setlength{\columnsep}{6.5mm}
\setlength\bibsep{1pt}

\makeatletter 
\newlength{\figrulesep} 
\setlength{\figrulesep}{0.5\textfloatsep} 

\newcommand{\topfigrule}{\vspace*{-1pt}%
\noindent{\color{cream}\rule[-\figrulesep]{\columnwidth}{1.5pt}} }

\newcommand{\botfigrule}{\vspace*{-2pt}%
\noindent{\color{cream}\rule[\figrulesep]{\columnwidth}{1.5pt}} }

\newcommand{\dblfigrule}{\vspace*{-1pt}%
\noindent{\color{cream}\rule[-\figrulesep]{\textwidth}{1.5pt}} }

\makeatother

\twocolumn[
  \begin{@twocolumnfalse}
{\hfill\raisebox{0pt}[0pt][0pt]{\includegraphics[height=55pt]{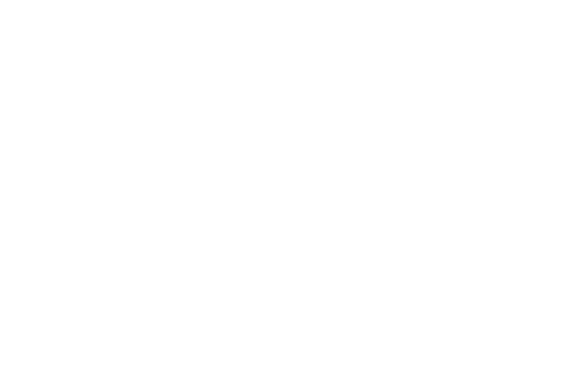}}\\[1ex]
}\par
\vspace{1em}
\sffamily
\begin{tabular}{m{0.5cm} p{17.0cm} }

 & \noindent\LARGE{\textbf{Improved SABRE hyperpolarisation using pulse sequences to reduce effective coupling}} \\
\vspace{0.3cm} & \vspace{0.3cm} \\

 & \noindent\large{Vitaly P. Kozinenko,\textit{$^{a}$} Bogdan A. Rodin,\textit{$^{a}$} James Eills,\textit{$^{b}$} Ilai Schwartz,\textit{$^{a}$} Stephan Knecht,$^{\ast}$\textit{$^{a}$} and Laurynas Dagys$^{\ast}$\textit{$^{c}$}} \\
 &\\

 & \noindent\normalsize{Hyperpolarisation using Signal Amplification By Reversible Exchange (SABRE) is a convenient method for high repeatability studies. The core of this technique is polarisation transfer to the target substrate during an on-going chemical exchange process. Typically, polarisation transfer is achieved as fast as possible. In this study we employ NMR sequences that on contrary slow down the polarisation transfer and yet demonstrate improved performance. Simulations confirm that such methods can lead to high polarisation yield in SABRE system that exhibit  higher magnetic inequivalence and lower chemical exchange rate.} \\

\end{tabular}

 \end{@twocolumnfalse} \vspace{0.6cm}

  ]

\renewcommand*\rmdefault{bch}\normalfont\upshape
\rmfamily
\section*{}
\vspace{-1cm}


\footnotetext{\textit{$^{a}$~NVision Imaging Technologies GmbH, Wolfgang-Paul Straße 2, 89081 Ulm, Germany.}}
\footnotetext{\textit{$^{b}$~Forschungszentrum Jülich,Wilhelm-Johnen-Straße,52428 Jülich, Germany.}}
\footnotetext{\textit{$^{c}$~Institute of Chemical Physics, Vilnius University, Saulėtekio av. 3, Vilnius LT10257, Lithuania.}}
\footnotetext{\textit{$^{\ast}$Correspondence: stephan@nvision-imaging.com; laurynas.dagys@ff.vu.lt}}



\section{Introduction}

Hyperpolarisation techniques offer a promising solution to the inherently low sensitivity of Nuclear Magnetic Resonance (NMR) spectroscopy. 
These methods enhance sensitivity by increasing nuclear spin-polarisation through various interactions with an external source of spin-order.\cite{bowersTransformationSymmetrizationOrder1986,eillsSpinHyperpolarizationModern2023,duckettApplicationParahydrogenInduced2012,fearSABREHyperpolarizedAnticancer2022,phamBiomolecularInteractionsStudied2023, tennantBenchtopNMRAnalysis2020,gemeinhardtDirect13CHyperpolarization2020,tomhonTemperatureCyclingEnables2022,demaissinVivoMetabolicImaging2023,pravdivtsevLIGHTSABREHyperpolarizes113CPyruvate2023,schmidt20Carbon13Polarization2023,myersDirectDetectionSABRESHEATH2025,erikssonImprovingSABREHyperpolarization2022,liSABREEnhancementOscillating2022,lindaleMultiaxisFieldsBoost2024,markelovAdiabaticApproachHeteronuclear2023,markelovHighfieldSABREPulse2024,kozinenkoSLICSABREMicroteslaFields2025,theisLIGHTSABREEnablesEfficient2014,truong15NHyperpolarizationReversible2015,theisMicroteslaSABREEnables2015,feketeRemarkableLevels15N2020,knechtEfficientConversionAntiphase2019,schmidt2013CHyperpolarization,schwartzRobustOpticalPolarization2018,tratzmillerParallelSelectiveNuclearspin2021}
A class of methods known as ParaHydrogen-Induced Polarisation (PHIP) derives signal enhancement from a chemical reaction with a specific spin isomer of hydrogen called parahydrogen.~\cite{bowersTransformationSymmetrizationOrder1986,eillsSpinHyperpolarizationModern2023,duckettApplicationParahydrogenInduced2012,fearSABREHyperpolarizedAnticancer2022,phamBiomolecularInteractionsStudied2023, tennantBenchtopNMRAnalysis2020,gemeinhardtDirect13CHyperpolarization2020,tomhonTemperatureCyclingEnables2022,demaissinVivoMetabolicImaging2023,pravdivtsevLIGHTSABREHyperpolarizes113CPyruvate2023,schmidt20Carbon13Polarization2023,myersDirectDetectionSABRESHEATH2025,erikssonImprovingSABREHyperpolarization2022,liSABREEnhancementOscillating2022,lindaleMultiaxisFieldsBoost2024,markelovAdiabaticApproachHeteronuclear2023,markelovHighfieldSABREPulse2024,kozinenkoSLICSABREMicroteslaFields2025,theisLIGHTSABREEnablesEfficient2014,truong15NHyperpolarizationReversible2015,theisMicroteslaSABREEnables2015,feketeRemarkableLevels15N2020,knechtEfficientConversionAntiphase2019,schmidt2013CHyperpolarization}
The protons of parahydrogen which are entangled in a nuclear singlet state deposit this state onto a target molecule by hydrogenation reaction. 
Then the singlet-order in the sample is converted into the magnetisation of the target spins using a variety of possible NMR sequences.
PHIP stands out as a cost-effective hyperpolarisation methodology with growing interest in the fields of medical imaging and metabolomics.~\cite{fearSABREHyperpolarizedAnticancer2022,eillsSpinHyperpolarizationModern2023,gemeinhardtDirect13CHyperpolarization2020,myersDirectDetectionSABRESHEATH2025,phamBiomolecularInteractionsStudied2023,tennantBenchtopNMRAnalysis2020}

A promising PHIP variant offering high repeatability and low cost is called Signal Amplification By Reversible Exchange (SABRE).\cite{eillsSpinHyperpolarizationModern2023,duckettApplicationParahydrogenInduced2012,fearSABREHyperpolarizedAnticancer2022,phamBiomolecularInteractionsStudied2023, tennantBenchtopNMRAnalysis2020,gemeinhardtDirect13CHyperpolarization2020,tomhonTemperatureCyclingEnables2022,demaissinVivoMetabolicImaging2023,pravdivtsevLIGHTSABREHyperpolarizes113CPyruvate2023,schmidt20Carbon13Polarization2023,myersDirectDetectionSABRESHEATH2025,erikssonImprovingSABREHyperpolarization2022,liSABREEnhancementOscillating2022,lindaleMultiaxisFieldsBoost2024,markelovAdiabaticApproachHeteronuclear2023,markelovHighfieldSABREPulse2024,kozinenkoSLICSABREMicroteslaFields2025,theisLIGHTSABREEnablesEfficient2014,truong15NHyperpolarizationReversible2015,theisMicroteslaSABREEnables2015,feketeRemarkableLevels15N2020,knechtEfficientConversionAntiphase2019,schmidt2013CHyperpolarization}
It utilizes a catalyst, typically iridium-based, that supports reversible formation of parahydrogen-substrate complex instead of irreversible hydrogenation.
The magnetisation of the target nuclei is converted from the singlet state of the attached parahydrogen using various polarisation transfer schemes.
Reversible chemical exchange then releases the substrate to the solution which gradually builds up polarisation in the free substrate pool.
The substrate is not consumed during the SABRE process, meaning that polarisation can be restored with a constant supply of parahydrogen gas.
This makes SABRE method a cost-effective hyperpolarisation method, finding applications in drug detection, \textit{ex-situ} metabolomics, and rapidly advancing towards \textit{ex-vivo} studies.~\cite{barskiyFeasibilityFormationKinetics2014,fearSABREHyperpolarizedAnticancer2022,phamBiomolecularInteractionsStudied2023, tennantBenchtopNMRAnalysis2020,gemeinhardtDirect13CHyperpolarization2020,demaissinVivoMetabolicImaging2023}

The SABRE mechanism stems from a fine-tuned interplay between chemical exchange and spin dynamics.
Finding suitable SABRE conditions often requires extensive and careful optimisation of both polarisation transfer and exchange rates as their matching determines the overall efficiency.
For example, too fast chemical exchange makes substrate-parahydrogen complex short-lived and does not allow polarisation transfer to effectively build up magnetisation in the substrate pool.
Therefore, one known strategy to increase SABRE efficiency is to slow down chemical exchange by lowering the temperature.~\cite{barskiyFeasibilityFormationKinetics2014,tomhonTemperatureCyclingEnables2022}
This is especially effective for hyperpolarisation of weak SABRE ligands like  \Cth-pyruvate where polarisation transfer is limited by the weak (sub-hertz) spin-spin coupling between the hydride protons and the \Cth site.~\cite{demaissinVivoMetabolicImaging2023,pravdivtsevLIGHTSABREHyperpolarizes113CPyruvate2023,schmidt20Carbon13Polarization2023,myersDirectDetectionSABRESHEATH2025}
Other substrates, particularly \Nlab-labelled heterocycle systems, form complexes with strong spin-spin coupling to the hydrides leading to much faster spin dynamics and  polarisation build-up. 
However, once the heteronuclear spin-spin coupling becomes larger than the proton homonuclear coupling in the hydride, it induces a strong magnetic inequivalence for the \Hone sites in the complex.
This considerably alters the dynamics of singlet-order to magnetisation conversion, rendering SABRE methodology sensitive to the optimisation of the applied NMR sequences.~\cite{bengsRobustTransformationSinglet2020,erikssonImprovingSABREHyperpolarization2022,liSABREEnhancementOscillating2022,lindaleMultiaxisFieldsBoost2024}

\begin{figure}[h]
\centering
  \includegraphics[width=0.45\textwidth]{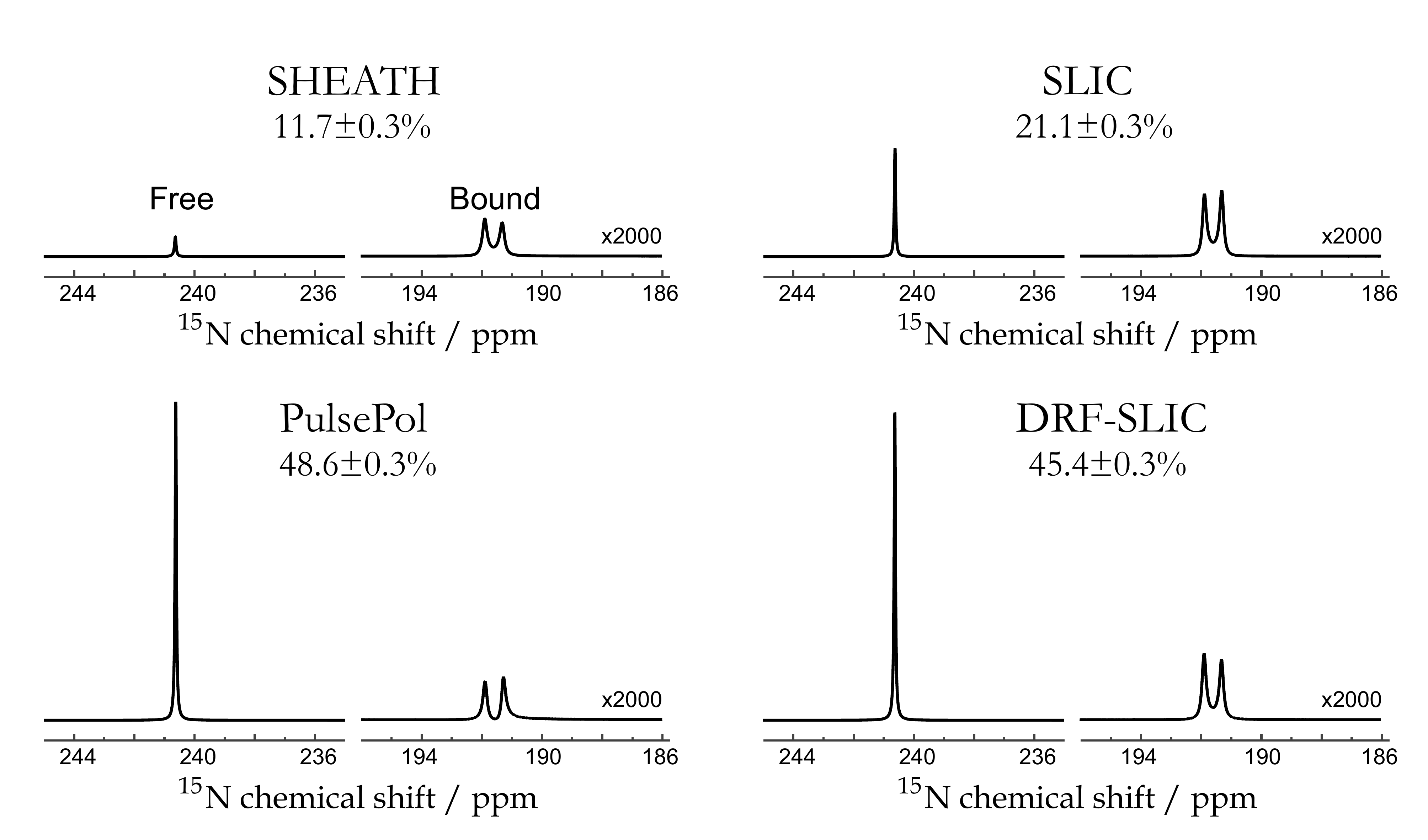}
  \caption{The \Nlab spectra of hyperpolarised \Nlab-acetonitrile recorded after four different SABRE protocols. Polarisation levels of the free substrate species are provided above each spectra. Deviation is estimated by repeating experiments four times. See subsection \ref{Sec:protocols} for more experimental details.}
  \label{fgr:teaser}
\end{figure}

In this work, we analyse two different NMR sequences - Double-Radio-Frequency Spin Lock-Induced Crossing (DRF-SLIC) and PulsePol, capable of effectively reducing heteronuclear spin-spin coupling during the polarisation transfer in the SABRE complex.~\cite{markelovAdiabaticApproachHeteronuclear2023,markelovHighfieldSABREPulse2024,kozinenkoSLICSABREMicroteslaFields2025,schwartzRobustOpticalPolarization2018,tratzmillerParallelSelectiveNuclearspin2021,sabbaSymmetrybasedSingletTriplet2022,korzeczekUnifiedPicturePolarization2024}
We compared these methods to the more established SABRE-SHEATH and SABRE-SLIC protocols and applied them to three different target substrates - \Nlab-acetonitrile, \Nlab-pyridine and metronidazole.~\cite{deviencePreparationNuclearSpin2013,theisLIGHTSABREEnablesEfficient2014,truong15NHyperpolarizationReversible2015,theisMicroteslaSABREEnables2015,feketeRemarkableLevels15N2020,knechtEfficientConversionAntiphase2019,schmidt2013CHyperpolarization}
Although DRF-SLIC and PulsePol differ in their design, they both achieved a similar striking improvement in hyperpolarisation of \Nlab-acetonitrile (Fig.~\ref{fgr:teaser}).
We hypothesise that this is linked to increased magnetic equivalence of the hydride and reduced polarisation transfer rate, approaching the rate of chemical exchange. 

This is confirmed by numerical simulations as well as by observed smaller and negative improvement in \Nlab-pyridine and metronidazole systems, respectively, which exhibit increasingly faster chemical exchange rate.
We therefore believe that methods like DRF-SLIC or PulsePol can out-perform other methods in many SABRE systems exhibiting similar dynamic range.
The additional degree of freedom in DRF-SLIC and PulsePol methods may potentially lead to more cost-effective SABRE applications further advancing hyperpolarisation towards routine NMR practice.

\section{Methods}

\subsection{Materials}
Three SABRE solutions were studied. Each solution was prepared by first dissolving 3~mM of \IrCat and the specified substrates in methanol-$\mathrm{d}_4$, followed by activation of the catalyst with hydrogen gas for a few minutes in an NMR pressure tube. For the \Nlab-acetonitrile, the solution contained 10~mM of \Nlab-acetonitrile as a substrate and 20~mM of pyridine as a co-substrate. The \Nlab-pyridine and metronidazole solutions contained 30~mM of the respective substrate.

The \Nlab-acetonitrile, \Nlab-pyridine, metronidazole, methanol-$\mathrm{d}_4$ where purchased from Sigma Aldrich while \IrCat was synthesised from [Ir(COD)Cl$_2$] and \IMes (IMes) using a protocol described in literature.~\cite{pavlikSynthesisSpectroscopicProperties2007,vazquez-serranoSearchNewHydrogenation2006}

Parahydrogen gas was produced using an ARS parahydrogen generator operating at a temperature of 22~K and stored in aluminium bottles at a pressure of 30~bar.
Parahydrogen was supplied to the SABRE solutions at 10~bar of pressure using a pressure regulator and electronically controlled valves.

\subsection{Equipment}

All experiments were performed using an automated system comprising a 9.4~T Bruker superconducting magnet, a Bruker Avance NEO NMR console, a TwinLeaf MS-1 $\mathrm{\mu}$-metal shield equipped with home-built coils, and a sample shuttling system integrated with a solenoid valve array managing the gas flow.
Hydrogen gas was supplied to the 5~mm pressure NMR tubes using 1/16" Teflon and 0.2~mm fused silica capillaries.
The entire system was controlled via a custom Python script running on a separate computer.
The static bias field during SABRE protocols was produced by solenoid coil powered by Rhode\&Schwartz HMP2030 power supply. The oscillating transverse field was generated via the analogue output of a National Instruments PXIe-6363 card and amplified by home-built audio amplifier. The waveforms for the low-field NMR sequences were calculated using the custom Python script and digitised at a 400~kHz sampling rate.

High-field \Nlab NMR spectra of 200~ppm width were collected with 64~k point density by exciting the SABRE-polarized samples with a single 17~$\mu$s 90\degr pulse.
The \Nlab enhancement factors were estimated by collecting thermally polarized \Nlab spectra of the SABRE samples with \Hone decoupling at room temperature using 300 transients with repetition delay of 200~s. 
Spectra were baseline corrected.

\subsection{SABRE experiments}
\label{Sec:protocols}

\begin{figure}[h]
\centering
  \includegraphics[width=0.45\textwidth]{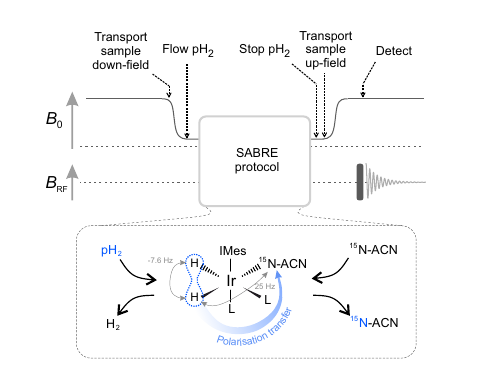}
  \caption{General scheme of the SABRE protocol. First, the sample is moved from the NMR magnet to a low magnetic field. Then, parahydrogen bubbling is initiated, followed by the polarisation transfer pulse sequence. Once sequence is finished, bubbling is stopped and sample is moved back to the magnet for recording of the NMR spectra. The inset shows the active SABRE complex between parahydrogen and \Nlab-acetonitrile. Relevant \textit{J}-couplings are indicated in grey.\cite{mewisStrategiesHyperpolarizationAcetonitrile2015} Letters "L" indicate other possible competing ligands, such as pyridine.}
  \label{fgr:SABRE_intro} 
\end{figure}

\begin{figure}[h]
\centering
  \includegraphics[width=0.45\textwidth]{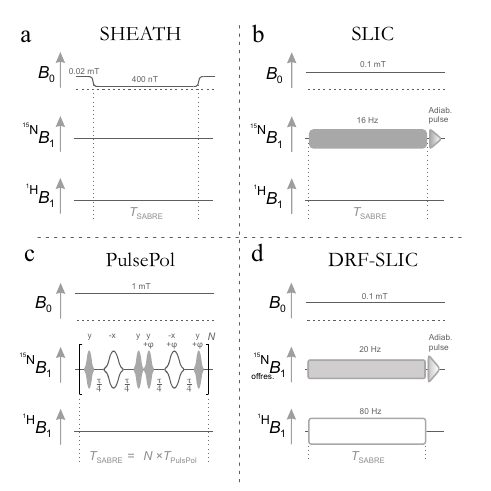}
  \caption{Low-field polarisation transfer sequences applied as depicted in Fig.~\ref{fgr:SABRE_intro} for hyperpolarisation of \Nlab compounds at low magnetic field. a) SHEATH involves ramping the magnetic field down to approximately 400~nT for the duration of the bubbling. b) Spin-Lock Induced Crossing (SLIC) is performed at fixed magnetic field at 98~$\mu$T by applying transverse field $B_1$ on \Nlab resonance with amplitude of $\sim$16~Hz. Final adiabatic pulse is applied to flip \Nlab magnetization along the bias field axis before sample transportation.  c) PulsePol method is performed at 1~mT bias field.  Filled and unfilled shapes represent 90\degr and 180\degr shaped pulses, respectively with Gaussian shape truncated at 10\%. Number of repetitions ($N$) is set to match bubbling time $T_{\mathrm{SABRE}}$. Duration $\tau$ and phase $\varphi$ was adjusted for optimal polarisation transfer. See more details in the text. d) Double-radio-frequency SLIC (DRF-SLIC) protocol uses on-resonance transverse field for \Hone and off-resonant field for \Nlab nuclei. The resonance mismatch is adjusted for optimal polarization transfer. Final adiabatic pulse is applied to flip \Nlab magnetisation along the bias field axis before sample transportation. }
  \label{fgr:sequences}
\end{figure}

The experiments in this study were performed following the procedure outlined in Fig.~\ref{fgr:SABRE_intro} using the automated experimental setup.
Activated SABRE solutions were transported to a low-field area where fresh parahydrogen was supplied via bubbling.
The bubbling period lasted 20~s during which SABRE protocols shown in figure~\ref{fgr:sequences} were executed.
After the samples were polarised, bubbling was ceased, and the samples were shuttled to the high field NMR magnet for data acquisition.

Experiments involving Spin-Lock Induced Crossing (SLIC) utilised a bias field of 98~$\mu$T and a transverse field resonant with \Nlab spins. The nutation frequency  of the SLIC pulse was optimised within the range of 0–30~Hz for each substrate. See Supplementary Information for more information.
At the end of the sequence adiabatic 90\degr pulse on \Nlab was used to align \Nlab magnetisation to the bias field before shuttling the sample.
The adiabatic pulses was performed by linearly ramping the pulse amplitude down and gradually decreasing pulse frequency by 50~Hz in 1~s.

SHEATH protocol involved switching magnetic field from 20~$\mu$T magnetic field to ultra-low magnetic field for the duration of the bubbling.
The SHEATH field was optimised within the range of 0-2000~nT.
After the bubbling with parahydrogen was finished, the magnetic field was switched back to 20~$\mu$T before the samples were shuttled back to the high field magnet.\cite{kiryutinCompleteMagneticField2018}

PulsePol NMR sequence was tested at a bias field of 1~mT with a 
Gaussian pulse peak amplitude of 600~Hz.
The shape ensured minimal Bloch-Siegert shift due to big pulse amplitudes and avoided the indirect excitation of \Hone spins due to short pulse durations.
The phase $\varphi$ and duration $\tau$ were experimentally optimised, using simulated values as an initial starting point.
The number of PulsePol cycles was adjusted to match the 20~s duration of the bubbling process.

Final experiments utilizing double-RF SLIC (DRF-SLIC) method were performed at 98~$\mu$T bias field using two excitation channels, one resonant with \Hone spins and second slightly off-resonant with \Nlab spins.
The transverse field amplitude for \Hone was fixed at 80~Hz while for \Nlab amplitude was fixed at 16-30~Hz with the resonance detuning of 60-100~Hz. 

A comprehensive summary of all experimental parameters is provided in the Supplementary Information (SI).

\section{Adjusting the effective coupling}
\label{sec:coupling}


Optimising SABRE efficiency requires a precise balance between coherent spin dynamics and dissociation of the SABRE-active complex (Fig.~\ref{fgr:SABRE_intro}). 
Generally, higher dissociation rates also suit higher optimal polarisation transfer rates but the precise correlation is non-trivial.
Cases exist where the highest efficiency is expected once population exchange ($A^*$) is comparable with the dissociation constant ($k_\mathrm{d}$), i.e.,  $A^* \approx k_d$.~\cite{barskiySimpleAnalyticalModel2015} 
These relationships constitute the first tuning constraints on pulse sequence design in SABRE.  
Within the framework of level anti-crossing (LAC) theory the population exchange rate is determined by the perturbation $V$  ($A^*=2V$) which breaks the degeneracy of two interacting states.~\cite{rodinRepresentationPopulationExchange2020}

The second constraint, particularly relevant for nitrogen-containing heterocycles such as \Nlab-acetonitrile, stems from the spin-coupling regime.   In SABRE complexes involving these molecules, the heteronuclear coupling $J_{\text{NH}}$ between the hydrides and the substrate \Nlab site often exceeds the homonuclear hydride coupling $J_{\text{HH}}$ leading to magnetic inequivalent spin system. Consequently, the standard LAC condition overlaps with others and leads to non-selective redistribution of populations to multiple spin states, interfering with the desired polarisation transfer. 
Therefore, reduced heteronuclear coupling leads to a narrow LAC conditions, restoring the isolated and efficient polarisation transfer.

Let us first consider the case of 3 spin-1/2 system under the influence of transverse oscillating field, slightly off-resonant for the \Nlab nuclei. Method with on-resonant field corresponds to a more known SLIC sequence.~\cite{deviencePreparationNuclearSpin2013} In the rotating tilted frame, the secular and perturbing Hamiltonians can be written as:~\cite{markelovAdiabaticApproachHeteronuclear2023}

\begin{equation}\label{eq:hamiltonian_slic}
\begin{gathered}
\hat{H} = \hat{H}_0 + \hat{V} \\
\hat{H}_0 = -\nu_0^\mathrm{H} \left(\hat{I}_\mathrm{H,z} + \hat{I}_\mathrm{{H',z}} \right) - \nu_{\text{eff}}^\mathrm{N} \hat{I}_\mathrm{N,z} + J_\mathrm{HH} \hat{\boldsymbol{I}}_\mathrm{H} \hat{\boldsymbol{I}}_\mathrm{H'} \\
+ J_\mathrm{NH}\frac{\cos{\theta}}{2}\,\left(\hat{I}_\mathrm{H,z} \hat{I}_\mathrm{N,z}+\hat{I}_\mathrm{H',z} \hat{I}_\mathrm{N,z}\right)
\\
\hat{V} = J_\mathrm{NH}\frac{\cos{\theta}}{2}\,\left(\hat{I}_\mathrm{H,z} \hat{I}_\mathrm{N,z}-\hat{I}_\mathrm{H',z} \hat{I}_\mathrm{N,z}\right)
\\ - J_\mathrm{NH} \sin{\theta} \, \hat{I}_\mathrm{H,z} \hat{I}_\mathrm{N,x} 
\end{gathered}
\end{equation}

\noindent where $\nu_{\text{eff}}^\mathrm{N} = \sqrt{(\nu_{\text{nut}}^\mathrm{N})^2 + (\Omega^\mathrm{N})^2}$ is the effective  frequency of \Nlab nucleus rotating around the field tilted with respect to the static field by the angle defined by $\tan{\theta}= \frac{\nu_{\text{nut}}^\mathrm{N}}{\Omega^\mathrm{N}}$. $I$ denotes the spin-operator, $\nu_0$ is the Larmor frequency of the respective nucleus, $\Omega$ is resonance offset, and $\nu_{\text{nut}}^\mathrm{N}$ is the nutation frequency of \Nlab nucleus. $J_\mathrm{HH}$ is the homonuclear coupling between the hydrides, $J_\mathrm{HN}$ is the heteronuclear coupling between one of the hydride protons and \Nlab site and $J_\mathrm{H'N}$ is assumed to be zero for simplicity.

The eigenbasis of the secular Hamiltonian $\hat{H}_0$ can be found by taking the tensor product of proton singlet-triplet basis and \Nlab Zeeman basis, forming the so-called $\mathbf{STZ}$ basis, expressed as $\{ |S\rangle, |T_0\rangle, |T_+\rangle, |T_-\rangle \}_H \otimes \{ |\alpha'\rangle, |\beta'\rangle \}_N$,  with $|S\rangle = \frac{1}{\sqrt{2}}(|\alpha\beta\rangle - |\beta\alpha\rangle)$, $|T_0\rangle = \frac{1}{\sqrt{2}}(|\alpha\beta\rangle + |\beta\alpha\rangle)$, $|T_+\rangle = |\alpha\alpha\rangle$, $|T_-\rangle = |\beta\beta\rangle$ and $\{ |\alpha'\rangle, |\beta'\rangle \}_N$  defined along the z-axis in the tilted frame.
For the purposes of analysing polarisation transfer during the SABRE experiment, one can isolate the states of interest - $\left\{ |S\alpha'\rangle, |S\beta'\rangle, |T_0\alpha'\rangle, |T_0\beta'\rangle \right\}$ and express the total Hamiltonian in the matrix form at the LAC condition ($\nu^\mathrm{N}_\mathrm{eff}=-J_\mathrm{HH}$):

\begin{equation}\label{eq:slic_matrix}
\resizebox{\columnwidth}{!}{
$
\hat{H} = 
\begin{array}{c|cccc} 
& \vert S \alpha' \rangle & \vert S \beta' \rangle & \vert T_0 \alpha' \rangle & \vert T_0 \beta' \rangle 
\\ \hline \langle S \alpha' \vert & -\frac{5}{4} J_\mathrm{HH} & 0 & \frac{1}{4} J_\mathrm{NH} \cos \theta & -\frac{1}{4} J_\mathrm{NH} \sin \theta 
\\ \langle S \beta' \vert & -- & -\frac{1}{4} J_\mathrm{NH} & -\frac{1}{4} J_\mathrm{NH} \sin \theta & -\frac{1}{4} J_\mathrm{NH} \cos \theta
\\ \langle T_0 \alpha' \vert & -- & -- & -\frac{1}{4} J_\mathrm{NH} & 0
\\ \langle T_0 \beta' \vert & -- & -- & -- & \frac{3}{4} J_\mathrm{HH}
\end{array} 
$}
\end{equation}

At this condition, the otherwise degenerate states $|S\beta'\rangle$ and $|T_0\alpha'\rangle$  are perturbed and undergo population exchange. 
Assuming fully populated proton singlet states, this exchange alone is sufficient to achieve fully polarised heteronucleus.
The effective transition rate is given by the off-diagonal element $2V = 2\langle S \beta' \vert\hat{H}_0\vert T_0 \alpha' \rangle= \frac{-1}{2} J_\mathrm{NH}\sin{\theta}$. Note that for the conventional SLIC method detuning parameter is $\Omega^\mathrm{N}=0$ and tilt angle $\theta$ equals $90^{\circ}$, ensuring maximum rate of $\frac{1}{2} J_\mathrm{NH}$. 
Reduction of the effective angle can be used as SABRE optimisation parameter for suitable matching to chemical exchange.
However, in three spin system with $|J_{\text{NH}}| > |J_{\text{HH}}|$, the perturbation connecting $|S\alpha'\rangle$ and $|T_0\beta'\rangle$  states may be relatively large compared to their energy difference ($\Delta E= 2J_\mathrm{HH}$).
This can cause additional population exchange reducing the polarisation efficiency which is consistent with our numerical SABRE simulation of the studied systems (see SI).

To solve this problem, we utilise double resonance DRF-SLIC, where an additional on-resonance transverse field is applied to the \Hone channel. With additional on-resonance irradiation for protons, the quantisation axis is flipped. It results in:

\begin{equation}\label{eq:hamiltonian_drf_slic}
\begin{gathered}
\hat{H} = \hat{H}_0 + \hat{V} \\
\hat{H}_0 = -\nu_{\text{nut}}^\mathrm{H} \left(\hat{I}_\mathrm{H,z} + \hat{I}_\mathrm{{H',z}} \right)- \nu_{\text{eff}}^\mathrm{N} \hat{I}_\mathrm{N,z} + J_\mathrm{HH} \hat{\boldsymbol{I}}_\mathrm{H}\hat{\boldsymbol{I}}_\mathrm{H'} \\
+ J_\mathrm{NH}\frac{\cos{\theta}}{2}\,\left(\hat{I}_\mathrm{H,x} \hat{I}_\mathrm{N,z}+\hat{I}_\mathrm{H',x} \hat{I}_\mathrm{N,z}\right) \\
\hat{V} = J_\mathrm{NH}\frac{\cos{\theta}}{2}\,\left(\hat{I}_\mathrm{H,x} \hat{I}_\mathrm{N,z}-\hat{I}_\mathrm{H',x} \hat{I}_\mathrm{N,z}\right) \\+ J_\mathrm{NH} \sin{\theta} \, \hat{I}_\mathrm{H,x} \hat{I}_\mathrm{N,x} 
\end{gathered}
\end{equation}

The resonance matching for DRF-SLIC is governed by the modified LAC condition: $\nu_{\text{eff}}^\mathrm{N} = \nu_{\text{nut}}^\mathrm{H} \pm J_\mathrm{HH}$. Here, the hydride spins are quantised along the $x$-axis of the rotating frame. We denote the resulting triplet states as $|T'\rangle$, while the singlet state $|S\rangle$ remains invariant under this rotation. The transition leading to heternulcear polarisation transfer is $|S \beta'\rangle \to |T'_-\alpha'\rangle$ if the state $|S \alpha'\rangle$ is unperturbed. The partial matrix representation of the Hamiltonian at the LAC condition is given by:

\begin{equation}\label{eq:drf_matrix}
\resizebox{\columnwidth}{!}{
$
\hat{H} = 
\begin{array}{c|cccc} 
& \vert S \alpha' \rangle & \vert S \beta' \rangle & \vert T'_- \alpha' \rangle & \vert T'_- \beta' \rangle 
\\ \hline \langle S \alpha' \vert & -\frac{1}{4} (5J_\mathrm{HH} + 2 \nu_{\text{nut}}^\mathrm{H}) & 0 & -\frac{1}{4 \sqrt{2}} J_\mathrm{NH} \cos \theta & \frac{1}{4 \sqrt{2}} J_\mathrm{NH} \sin \theta 
\\ \langle S \beta' \vert & -- & -\frac{1}{4} (J_\mathrm{NH} - 2\nu_{\text{nut}}^\mathrm{H}) & \frac{1}{4\sqrt{2}} J_\mathrm{NH} \sin \theta & \frac{1}{4\sqrt{2}} J_\mathrm{NH} \cos \theta
\\ \langle T'_0 \alpha' \vert & -- & -- & -\frac{1}{4}(J_\mathrm{NH} - 2\nu_{\text{nut}}^\mathrm{H}) & 0
\\ \langle T'_0 \beta' \vert & -- & -- & -- & -\frac{3}{4}(-J_\mathrm{NH} - 2\nu_{\text{nut}}^\mathrm{H})
\end{array}
$}
\end{equation}

\noindent where the rate of transition is $2 V = \frac{2}{4 \sqrt{2}}J_\mathrm{NH} \sin{\theta}$. Here, the key parameter is $\nu_{\text{nut}}^\mathrm{H}$ as it allows shifting the energy of $|S \alpha'\rangle$ state and isolate it from all the other states. Numerical simulations support these considerations, demonstrating that DRF-SLIC achieves consistently higher polarisation compared to single-frequency SLIC in the slow and intermediate exchange regimes. In the fast-exchange limit ($k_\mathrm{d} > 100~\text{s}^{-1}$), both methods exhibit comparable efficiency.

\textbf{PulsePol} represents an alternative approach to singlet-to-magnetisation conversion. It relies on applying a pulse train to construct an effective interaction-frame Hamiltonian, so the evolution of one of its harmonics matches the evolution frequency of the coupled states. Detailed analysis of this sequence can be found elsewhere.~\cite{schwartzRobustOpticalPolarization2018}

In the context of the present study, the primary advantage of PulsePol is that, depending on the selected resonance condition, the sequence can effectively rescale the heteronuclear scalar coupling. This provides a precise mechanism for matching the polarisation transfer rate to the catalyst dissociation rate ($k_\mathrm{d}$), analogous to the control achieved via DRF-SLIC, but with potentially greater robustness to experimental offsets.

\begin{figure}[h]
\centering
  \includegraphics[width=0.48\textwidth]{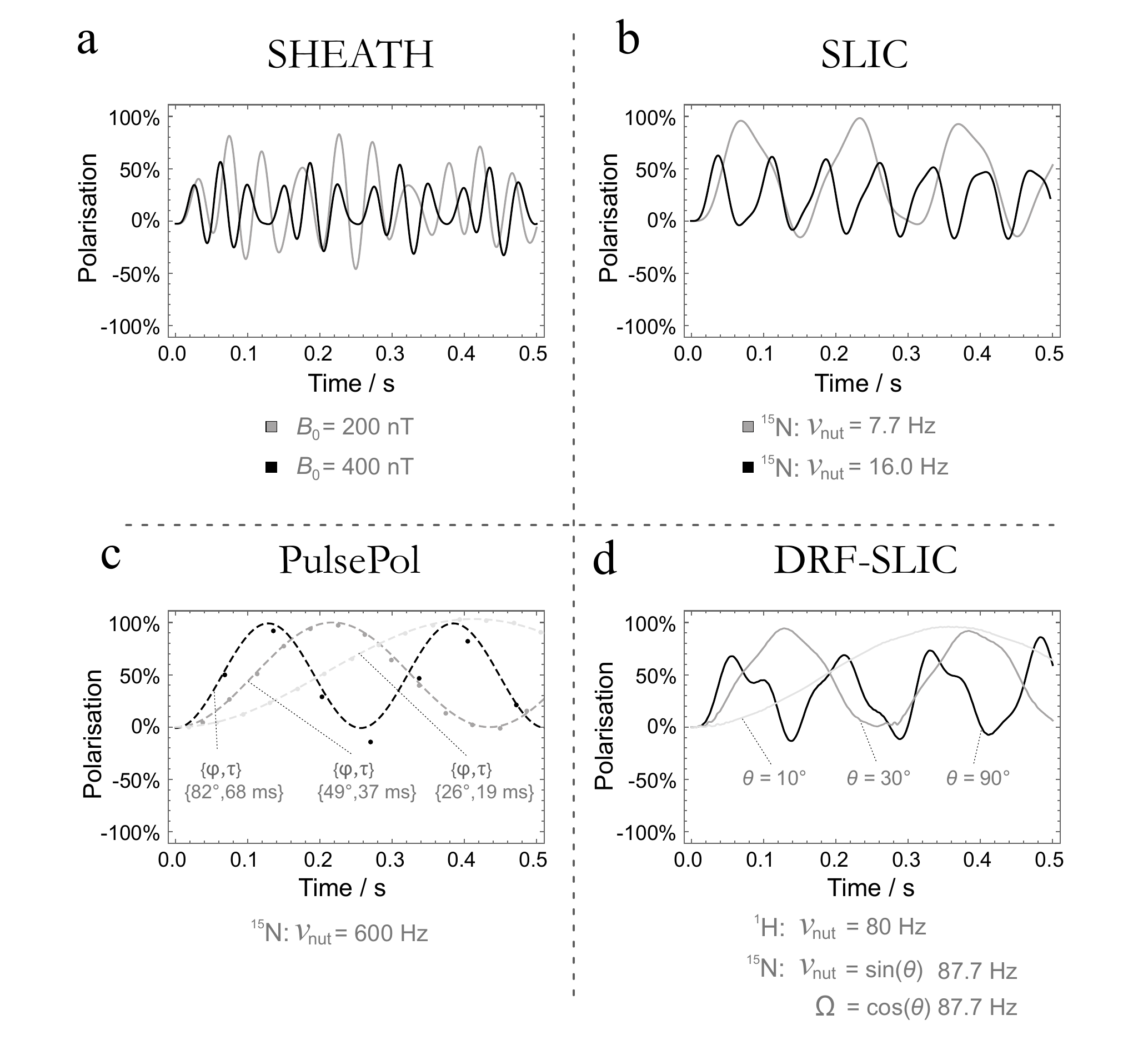}
  \caption{Numerical simulation of different singlet-to-magnetisation transfer methods in the active \Nlab-acetonitrile SABRE complex (depicted in Fig.~\ref{fgr:SABRE_intro}. a) Simulation of SHEATH transfer at 200~nT and 400~nT magnetic fields are given in grey and black, respectively.  b) Polarisation transfer using \Nlab on-resonance field depicted in grey and black for amplitudes 7.7~Hz and 16~Hz. c) Numerical simulation of \Nlab magnetisation for each PulsePol cycle at different phase $\varphi$ and duration $\tau$ settings and pulse amplitude set to 600~Hz. Sinusoidal curve is provided as a guide. d) Simulation of polarisation transfer trajectories during DRF-SLIC method are calculated according the mathematical relations given below. Simulations are given for three different effective angles $\theta$.}
  \label{fgr:Polcurves} 
\end{figure}


\begin{figure}[h]
\centering
  \includegraphics[width=0.35\textwidth]{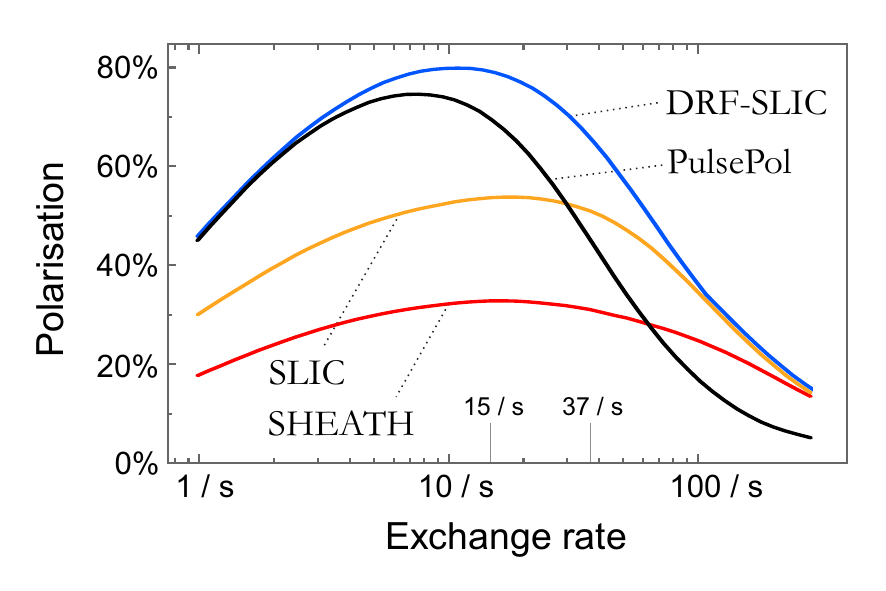}
  \caption{Numerically optimised SABRE performance for hyperpolarisation of \Nlab-acetonitrile as a function of exchange rate using different polarisation transfer methods: SHEATH (red line), SLIC (orange line), PulsePol (black line) and DRF-SLIC (blue line). Simulation assumes a rapid parahydrogen exchange and slow dissociation of a single \Nlab-molecule. Polarisation is evaluated after 20~s of its build-up.}
  \label{fgr:simulation} 
\end{figure}

\section{Results and discussion}

\begin{figure}[h]
\centering
  \includegraphics[width=0.35\textwidth]{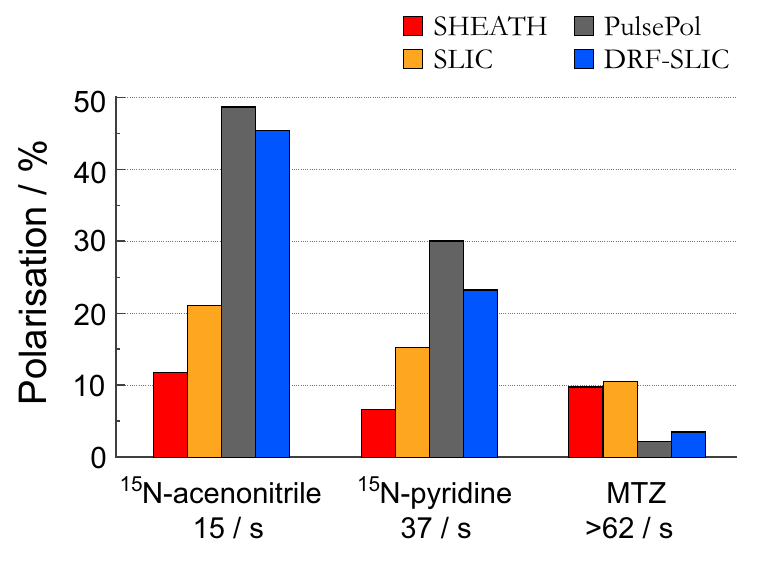}
  \caption{\Nlab polarisation levels experimentally achieved using different optimised SABRE protocols shown in Fig.~\ref{fgr:sequences} for three different compounds: \Nlab-acetonitrile,\Nlab-pyridine, metronidazole (MTZ). Experimentally determined dissociation rates at room temperature are given below. Metronidazole dissociation rate was estimated at 10\degC and thus is expected to be larger at room temperature.}
  \label{fgr:results}
\end{figure}

The \Nlab polarisation levels achieved after experimental optimisation of each SABRE protocol for the three target molecules are presented in Fig.~\ref{fgr:results}. 
We observe that results with each target molecule vary considerably depending on the applied SABRE method.
The most striking difference was observed on \Nlab-acetonitrile where DRF-SLIC and PulsePol led to 45\% and 49\%, respectively, compared to more established SLIC and SHEATH methods (21\% and 12\%, respectively).
This improvement is also qualitatively predicted by the numerical simulation (Fig.~\ref{fgr:simulation}) although simulation suggests that DRF-SLIC should always outperform.
As the PulsePol sequence offers higher robustness to resonance offsets we link the reduced efficiency of DRF-SLIC to experimental imperfections.

Since, as detailed in Section \ref{sec:coupling}, both PulsePol and DRF-SLIC allow scaling of the effective heteronuclear coupling in the SABRE complex, the observed higher performance (compared to SHEATH and SLIC) could be associated with two important factors.
First, the reduced polarisation rate can be more suited for the slow chemical exchange rather the rapid one.
Second, weaker effective heteronuclear coupling ensures strong-coupling regime for the two hydride \Hone spins which in turn leads to more isolated polarisation build-up (see Fig.~\ref{fgr:Polcurves}).

However, reducing the effective coupling is not a good strategy in systems with exceedingly high dissociation rates.
This is displayed by the less pronounced enhancement in polarisation for \Nlab-pyridine and the negative improvement (relative to SHEATH and SLIC) observed for metronidazole.

Our simulation suggests that at high exchange rates optimal conditions for DRF-SLIC converge back towards conventional SLIC where effective heteronuclear $J$-coupling is maximized and not reduced.
At dissociation rates above 100 1/s DRF-SLIC should not yield better results for SABRE systems with similar topology.
PulsePol method exhibits a tighter performance limit at exchange rates of around 40 1/s and above.
The likely reason is that PulsePol sequence relies on the pulse cycle completion for polarisation transfer.
Thus, if SABRE complex lifetime becomes comparable or shorter than the single PulsePol cycle, the polarisation can not be successfully transferred from the hydride to the \Nlab site causing a steep loss in performance.
The polarisation yield using PulsePol may be recovered by lowering the sample temperature as shown in the SI.
Therefore, we expect that the overall best SABRE performance can be achieved by adjusting both the effective heteronuclear coupling with a suitable sequence and the chemical exchange dynamics with a chemical and physical modification of the system.

\section*{Conclusions}

In this study, we investigated three SABRE systems possessing similar spin-spin coupling topologies but distinct chemical exchange kinetics.
These systems and the achieved \Nlab polarisation were compared by utilizing four different SABRE polarisation transfer protocols.
The highest performance was achieved using PulsePol and DRF-SLIC protocols with \Nlab-acetonitrile approaching 50\% of \Nlab polarisation. 
Compared to SABRE-SHEATH and SABRE-SLIC, these methods offer the adjustment of the effective heteronuclear coupling.
We hypothesize that such reduction of the driving interaction can increase the overall SABRE performance due to improved matching with chemical exchange and more isolated spin dynamics.
Evidently this strategy is not optimal in systems displaying fast chemical exchange and may not be beneficial for systems exhibiting high magnetic equivalence (e.g. SABRE of \Cth-pyruvate). 
However, a considerable variety of \Nlab-labelled SABRE substrates lie within these extremes and can lead to better results using sequence like PulsePol or DRF-SLIC.
These methods should further complement cases where chemical dynamics are slowed down by the reduction of temperature or addition of co-ligands.
Therefore, the fine tuning of the spin dynamics in SABRE can be an important aspect in improving efficiency of modern applications of hyperpolarisation.

\section*{Conflicts of interest}

The authors declare the following competing financial interests: V.P.K., B.A.R., I.S., S.K. and L.D. are or were employed by NVision Imaging Technologies GmbH. All other authors declare that they have no competing interests.

\section*{Data availability}

All data needed to evaluate the conclusions in the paper are present in the paper and the supplementary information. The experimental data, simulation scripts used in this manuscript are available at \note{[repository link to be added]}.

\section*{Acknowledgements}
L.D. would like to acknowledge Marius Jakulis Jason foundation for their support and funding received by the Research Council of Lithuania (grant number S-MIP-25-24).





\bibliography{refs} 

@article{barskiyFeasibilityFormationKinetics2014,
  title = {The {{Feasibility}} of {{Formation}} and {{Kinetics}} of {{NMR Signal Amplification}} by {{Reversible Exchange}} ({{SABRE}}) at {{High Magnetic Field}} (9.4 {{T}})},
  author = {Barskiy, Danila A. and Kovtunov, Kirill V. and Koptyug, Igor V. and He, Ping and Groome, Kirsten A. and Best, Quinn A. and Shi, Fan and Goodson, Boyd M. and Shchepin, Roman V. and Coffey, Aaron M. and Waddell, Kevin W. and Chekmenev, Eduard Y.},
  year = 2014,
  month = mar,
  journal = {Journal of the American Chemical Society},
  volume = {136},
  number = {9},
  pages = {3322--3325},
  publisher = {American Chemical Society},
  issn = {0002-7863},
  doi = {10.1021/ja501052p},
  urldate = {2025-05-23}
}

@article{korzeczekUnifiedPicturePolarization2024,
  title = {Towards a Unified Picture of Polarization Transfer --- Pulsed {{DNP}} and Chemically Equivalent {{PHIP}}},
  author = {Korzeczek, Martin C. and Dagys, Laurynas and M{\"u}ller, Christoph and Tratzmiller, Benedikt and Salhov, Alon and Eichhorn, Tim and Scheuer, Jochen and Knecht, Stephan and Plenio, Martin B. and Schwartz, Ilai},
  year = 2024,
  month = may,
  journal = {Journal of Magnetic Resonance},
  volume = {362},
  pages = {107671},
  issn = {1090-7807},
  doi = {10.1016/j.jmr.2024.107671},
  urldate = {2025-05-26}
}

@article{kozinenkoSLICSABREMicroteslaFields2025,
  title = {{{SLIC-SABRE}} at {{Microtesla Fields Enables High Levels}} of {{Nuclear Spin Polarization Without Magnetic Shielding}}},
  author = {Kozinenko, Vitaly P. and Kiryutin, Alexey S. and Yurkovskaya, Alexandra V.},
  year = 2025,
  journal = {Chemistry--Methods},
  volume = {5},
  number = {5},
  pages = {e202400060},
  issn = {2628-9725},
  doi = {10.1002/cmtd.202400060},
  urldate = {2025-05-24},
  copyright = {\copyright{} 2025 The Author(s). Chemistry - Methods published by Chemistry Europe and Wiley-VCH GmbH},
  langid = {english}
}

@article{markelovAdiabaticApproachHeteronuclear2023,
  title = {Adiabatic Approach for Heteronuclear {{SABRE}} Hyperpolarization at High Magnetic Field},
  author = {Markelov, Danil A. and Kozinenko, Vitaly P. and Yurkovskaya, Alexandra V. and Ivanov, Konstantin L.},
  year = 2023,
  month = dec,
  journal = {Journal of Magnetic Resonance Open},
  volume = {16--17},
  pages = {100139},
  issn = {2666-4410},
  doi = {10.1016/j.jmro.2023.100139},
  urldate = {2025-06-10}
}

@article{pavlikSynthesisSpectroscopicProperties2007,
  title = {Synthesis and Spectroscopic Properties of Isomeric Trideuterio- and Tetradeuterio Pyridines},
  author = {Pavlik, James W. and Laohhasurayotin, Somchoke},
  year = 2007,
  journal = {Journal of Heterocyclic Chemistry},
  volume = {44},
  number = {6},
  pages = {1485--1492},
  issn = {1943-5193},
  doi = {10.1002/jhet.5570440637},
  urldate = {2025-06-18},
  copyright = {Copyright \copyright{} 2007 Journal of Heterocyclic Chemistry},
  langid = {english}
}

@article{pravdivtsevLIGHTSABREHyperpolarizes113CPyruvate2023,
  title = {{{LIGHT-SABRE Hyperpolarizes}} 1-{{13C-Pyruvate Continuously}} without {{Magnetic Field Cycling}}},
  author = {Pravdivtsev, Andrey N. and Buckenmaier, Kai and Kempf, Nicolas and Stevanato, Gabriele and Scheffler, Klaus and Engelmann, Joern and Plaumann, Markus and Koerber, Rainer and H{\"o}vener, Jan-Bernd and Theis, Thomas},
  year = 2023,
  month = apr,
  journal = {The Journal of Physical Chemistry C},
  volume = {127},
  number = {14},
  pages = {6744--6753},
  publisher = {American Chemical Society},
  issn = {1932-7447},
  doi = {10.1021/acs.jpcc.3c01128},
  urldate = {2025-06-05}
}

@article{schmidt20Carbon13Polarization2023,
  title = {Over 20\% {{Carbon-13 Polarization}} of {{Perdeuterated Pyruvate Using Reversible Exchange}} with {{Parahydrogen}} and {{Spin-Lock Induced Crossing}} at 50 {{$\mu$T}}},
  author = {Schmidt, Andreas B. and Eills, James and Dagys, Laurynas and Gierse, Martin and Keim, Michael and Lucas, Sebastian and Bock, Michael and Schwartz, Ilai and Zaitsev, Maxim and Chekmenev, Eduard Y. and Knecht, Stephan},
  year = 2023,
  month = jun,
  journal = {The Journal of Physical Chemistry Letters},
  volume = {14},
  number = {23},
  pages = {5305--5309},
  publisher = {American Chemical Society},
  doi = {10.1021/acs.jpclett.3c00707},
  urldate = {2025-05-23}
}

@article{vazquez-serranoSearchNewHydrogenation2006,
  title = {The Search for New Hydrogenation Catalyst Motifs Based on {{N-heterocyclic}} Carbene Ligands},
  author = {{Vazquez-Serrano}, Leslie D. and Owens, Bridget T. and Buriak, Jillian M.},
  year = 2006,
  month = jun,
  journal = {Inorganica Chimica Acta},
  series = {Protagonists in {{Chemistry}}: {{Brian James}}},
  volume = {359},
  number = {9},
  pages = {2786--2797},
  issn = {0020-1693},
  doi = {10.1016/j.ica.2005.10.049},
  urldate = {2025-06-18}
}

@article{bengsRobustTransformationSinglet2020,
  title = {Robust Transformation of Singlet Order into Heteronuclear Magnetisation over an Extended Coupling Range},
  author = {Bengs, Christian and Dagys, Laurynas and Levitt, Malcolm H.},
  year = 2020,
  month = dec,
  journal = {Journal of Magnetic Resonance},
  volume = {321},
  pages = {106850},
  issn = {1090-7807},
  doi = {10.1016/j.jmr.2020.106850},
  urldate = {2026-01-26}
}

@article{bowersTransformationSymmetrizationOrder1986,
  title = {Transformation of {{Symmetrization Order}} to {{Nuclear-Spin Magnetization}} by {{Chemical Reaction}} and {{Nuclear Magnetic Resonance}}},
  author = {Bowers, C. Russell and Weitekamp, Daniel P.},
  year = 1986,
  month = nov,
  journal = {Physical Review Letters},
  volume = {57},
  number = {21},
  pages = {2645--2648},
  publisher = {American Physical Society},
  doi = {10.1103/PhysRevLett.57.2645},
  urldate = {2026-01-26}
}

@article{demaissinVivoMetabolicImaging2023,
  title = {In {{Vivo Metabolic Imaging}} of [1-{{13C}}]{{Pyruvate-d3 Hyperpolarized By Reversible Exchange With Parahydrogen}}},
  author = {{de Maissin}, Henri and Gro{\ss}, Philipp R. and Mohiuddin, Obaid and Weigt, Moritz and Nagel, Luca and Herzog, Marvin and Wang, Zirun and Willing, Robert and Reichardt, Wilfried and Pichotka, Martin and He{\ss}, Lisa and Reinheckel, Thomas and Jessen, Henning J. and Zeiser, Robert and Bock, Michael and {von Elverfeldt}, Dominik and Zaitsev, Maxim and Korchak, Sergey and Gl{\"o}ggler, Stefan and H{\"o}vener, Jan-Bernd and Chekmenev, Eduard Y. and Schilling, Franz and Knecht, Stephan and Schmidt, Andreas B.},
  year = 2023,
  journal = {Angewandte Chemie International Edition},
  volume = {62},
  number = {36},
  pages = {e202306654},
  issn = {1521-3773},
  doi = {10.1002/anie.202306654},
  urldate = {2026-01-26},
  copyright = {\copyright{} 2023 The Authors. Angewandte Chemie International Edition published by Wiley-VCH GmbH},
  langid = {english}
}

@article{deviencePreparationNuclearSpin2013,
  title = {Preparation of {{Nuclear Spin Singlet States Using Spin-Lock Induced Crossing}}},
  author = {DeVience, Stephen J. and Walsworth, Ronald L. and Rosen, Matthew S.},
  year = 2013,
  month = oct,
  journal = {Physical Review Letters},
  volume = {111},
  number = {17},
  pages = {173002},
  publisher = {American Physical Society},
  doi = {10.1103/PhysRevLett.111.173002},
  urldate = {2026-01-26}
}

@article{duckettApplicationParahydrogenInduced2012,
  title = {Application of {{Parahydrogen Induced Polarization Techniques}} in {{NMR Spectroscopy}} and {{Imaging}}},
  author = {Duckett, Simon B. and Mewis, Ryan E.},
  year = 2012,
  month = aug,
  journal = {Accounts of Chemical Research},
  volume = {45},
  number = {8},
  pages = {1247--1257},
  publisher = {American Chemical Society},
  issn = {0001-4842},
  doi = {10.1021/ar2003094},
  urldate = {2026-01-26}
}

@article{eillsSpinHyperpolarizationModern2023,
  title = {Spin {{Hyperpolarization}} in {{Modern Magnetic Resonance}}},
  author = {Eills, James and Budker, Dmitry and Cavagnero, Silvia and Chekmenev, Eduard Y. and Elliott, Stuart J. and Jannin, Sami and Lesage, Anne and Matysik, J{\"o}rg and Meersmann, Thomas and Prisner, Thomas and Reimer, Jeffrey A. and Yang, Hanming and Koptyug, Igor V.},
  year = 2023,
  month = feb,
  journal = {Chemical Reviews},
  volume = {123},
  number = {4},
  pages = {1417--1551},
  publisher = {American Chemical Society},
  issn = {0009-2665},
  doi = {10.1021/acs.chemrev.2c00534},
  urldate = {2026-01-26}
}

@article{erikssonImprovingSABREHyperpolarization2022,
  title = {Improving {{SABRE}} Hyperpolarization with Highly Nonintuitive Pulse Sequences: {{Moving}} beyond Avoided Crossings to Describe Dynamics},
  shorttitle = {Improving {{SABRE}} Hyperpolarization with Highly Nonintuitive Pulse Sequences},
  author = {Eriksson, Shannon L. and Lindale, Jacob R. and Li, Xiaoqing and Warren, Warren S.},
  year = 2022,
  month = mar,
  journal = {Science Advances},
  volume = {8},
  number = {11},
  pages = {eabl3708},
  publisher = {American Association for the Advancement of Science},
  doi = {10.1126/sciadv.abl3708},
  urldate = {2026-01-26}
}

@article{fearSABREHyperpolarizedAnticancer2022,
  title = {{{SABRE}} Hyperpolarized Anticancer Agents for Use in {{1H MRI}}},
  author = {Fear, Elizabeth J. and Kennerley, Aneurin J. and Rayner, Peter J. and Norcott, Philip and Roy, Soumya S. and Duckett, Simon B.},
  year = 2022,
  month = jul,
  journal = {Magnetic Resonance in Medicine},
  volume = {88},
  number = {1},
  pages = {11--27},
  issn = {0740-3194},
  doi = {10.1002/mrm.29166},
  urldate = {2026-01-26},
  pmcid = {PMC9310590},
  pmid = {35253267}
}

@article{feketeRemarkableLevels15N2020,
  title = {Remarkable {{Levels}} of {{15N Polarization Delivered}} through {{SABRE}} into {{Unlabeled Pyridine}}, {{Pyrazine}}, or {{Metronidazole Enable Single Scan NMR Quantification}} at the {{mM Level}}},
  author = {Fekete, Marianna and Ahwal, Fadi and Duckett, Simon B.},
  year = 2020,
  month = jun,
  journal = {The Journal of Physical Chemistry B},
  volume = {124},
  number = {22},
  pages = {4573--4580},
  publisher = {American Chemical Society},
  issn = {1520-6106},
  doi = {10.1021/acs.jpcb.0c02583},
  urldate = {2026-01-26}
}

@article{gemeinhardtDirect13CHyperpolarization2020,
  title = {``{{Direct}}'' {{13C Hyperpolarization}} of {{13C-Acetate}} by {{MicroTesla NMR Signal Amplification}} by {{Reversible Exchange}} ({{SABRE}})},
  author = {Gemeinhardt, Max E. and Limbach, Miranda N. and Gebhardt, Thomas R. and Eriksson, Clark W. and Eriksson, Shannon L. and Lindale, Jacob R. and Goodson, Elysia A. and Warren, Warren S. and Chekmenev, Eduard Y. and Goodson, Boyd M.},
  year = 2020,
  journal = {Angewandte Chemie International Edition},
  volume = {59},
  number = {1},
  pages = {418--423},
  issn = {1521-3773},
  doi = {10.1002/anie.201910506},
  urldate = {2026-01-26},
  copyright = {\copyright{} 2020 Wiley-VCH Verlag GmbH \& Co. KGaA, Weinheim},
  langid = {english}
}

@article{knechtEfficientConversionAntiphase2019,
  title = {Efficient Conversion of Anti-Phase Spin Order of Protons into {{15N}} Magnetisation Using {{SLIC-SABRE}}},
  author = {Knecht, Stephan and Kiryutin, Alexey S. and Yurkovskaya, Alexandra V. and Ivanov, Konstantin L.},
  year = 2019,
  month = oct,
  journal = {Molecular Physics},
  volume = {117},
  number = {19},
  pages = {2762--2771},
  publisher = {Taylor \& Francis},
  issn = {0026-8976},
  doi = {10.1080/00268976.2018.1515999},
  urldate = {2026-01-26}
}

@article{lindaleMultiaxisFieldsBoost2024,
  title = {Multi-Axis Fields Boost {{SABRE}} Hyperpolarization},
  author = {Lindale, Jacob R. and Smith, Loren L. and Mammen, Mathew W. and Eriksson, Shannon L. and Everhart, Lucas M. and Warren, Warren S.},
  year = 2024,
  month = apr,
  journal = {Proceedings of the National Academy of Sciences},
  volume = {121},
  number = {14},
  pages = {e2400066121},
  publisher = {Proceedings of the National Academy of Sciences},
  doi = {10.1073/pnas.2400066121},
  urldate = {2026-01-26}
}

@article{liSABREEnhancementOscillating2022,
  title = {{{SABRE}} Enhancement with Oscillating Pulse Sequences},
  author = {Li, Xiaoqing and Lindale, Jacob R. and Eriksson, Shannon L. and Warren, Warren S.},
  year = 2022,
  month = jul,
  journal = {Physical Chemistry Chemical Physics},
  volume = {24},
  number = {27},
  pages = {16462--16470},
  publisher = {The Royal Society of Chemistry},
  issn = {1463-9084},
  doi = {10.1039/D2CP00899H},
  urldate = {2026-01-26},
  langid = {english}
}

@article{markelovHighfieldSABREPulse2024,
  title = {High-Field {{SABRE}} Pulse Sequence Design for Chemically Non-Equivalent Spin Systems},
  author = {Markelov, Danil A. and Kozinenko, Vitaly P. and Kiryutin, Alexey S. and Yurkovskaya, Alexandra V.},
  year = 2024,
  month = dec,
  journal = {The Journal of Chemical Physics},
  volume = {161},
  number = {21},
  pages = {214203},
  issn = {0021-9606},
  doi = {10.1063/5.0236841},
  urldate = {2026-01-26}
}

@article{myersDirectDetectionSABRESHEATH2025,
  title = {Direct Detection of {{SABRE-SHEATH}} Hyperpolarization and Spin-Lattice Relaxation of [1-{{13C}}]Pyruvate},
  author = {Myers, John Z. and Plaumann, Markus and Buckenmaier, Kai and Pravdivtsev, Andrey N. and K{\"o}rber, Rainer},
  year = 2025,
  month = dec,
  journal = {Communications Chemistry},
  volume = {9},
  number = {1},
  pages = {44},
  publisher = {Nature Publishing Group},
  issn = {2399-3669},
  doi = {10.1038/s42004-025-01851-1},
  urldate = {2026-01-26},
  copyright = {2025 The Author(s)},
  langid = {english}
}

@article{phamBiomolecularInteractionsStudied2023,
  title = {Biomolecular Interactions Studied by Low-Field {{NMR}} Using {{SABRE}} Hyperpolarization},
  author = {Pham, Pierce and Hilty, Christian},
  year = 2023,
  journal = {Chemical Science},
  volume = {14},
  number = {37},
  pages = {10258--10263},
  publisher = {Royal Society of Chemistry},
  doi = {10.1039/D3SC02365F},
  urldate = {2026-01-26},
  langid = {english}
}

@article{sabbaSymmetrybasedSingletTriplet2022,
  title = {Symmetry-Based Singlet--Triplet Excitation in Solution Nuclear Magnetic Resonance},
  author = {Sabba, Mohamed and Wili, Nino and Bengs, Christian and Whipham, James W. and Brown, Lynda J. and Levitt, Malcolm H.},
  year = 2022,
  month = oct,
  journal = {The Journal of Chemical Physics},
  volume = {157},
  number = {13},
  pages = {134302},
  issn = {0021-9606},
  doi = {10.1063/5.0103122},
  urldate = {2026-01-26}
}

@article{schmidt2013CHyperpolarization,
  title = {Over 20\% {{13C Hyperpolarization}} for {{Pyruvate Using Deuteration}} and {{Rapid SLIC-SABRE}} in {{Mictrotesla Fields}}},
  author = {Schmidt, Andreas B. and Eills, James and Dagys, Laurynas and Gierse, Martin and Keim, Michael and Lucas, Sebastian and Bock, Michael and Schwartz, Ilai and Zaitsev, Maxim and Chekmenev, Eduard Y. and Knecht, Stephan},
  journal = {ChemRxiv},
  volume = {2023},
  number = {0224},
  pages = {11790--11799},
  publisher = {ChemRxiv},
  doi = {10.26434/chemrxiv-2023-ggvn4},
  urldate = {2026-01-26}
}

@article{schwartzRobustOpticalPolarization2018,
  title = {Robust Optical Polarization of Nuclear Spin Baths Using {{Hamiltonian}} Engineering of Nitrogen-Vacancy Center Quantum Dynamics},
  author = {Schwartz, Ilai and Scheuer, Jochen and Tratzmiller, Benedikt and M{\"u}ller, Samuel and Chen, Qiong and Dhand, Ish and Wang, Zhen-Yu and M{\"u}ller, Christoph and Naydenov, Boris and Jelezko, Fedor and Plenio, Martin B.},
  year = 2018,
  month = aug,
  journal = {Science Advances},
  volume = {4},
  number = {8},
  pages = {eaat8978},
  publisher = {American Association for the Advancement of Science},
  doi = {10.1126/sciadv.aat8978},
  urldate = {2026-01-26}
}

@article{tennantBenchtopNMRAnalysis2020,
  title = {Benchtop {{NMR}} Analysis of Piperazine-Based Drugs Hyperpolarised by {{SABRE}}},
  author = {Tennant, Thomas and Hulme, Matthew C. and Robertson, Thomas B.R. and Sutcliffe, Oliver B. and Mewis, Ryan E.},
  year = 2020,
  journal = {Magnetic Resonance in Chemistry},
  volume = {58},
  number = {12},
  pages = {1151--1159},
  issn = {1097-458X},
  doi = {10.1002/mrc.4999},
  urldate = {2026-01-26},
  copyright = {\copyright{} 2020 John Wiley \& Sons, Ltd.},
  langid = {english}
}

@article{theisLIGHTSABREEnablesEfficient2014,
  title = {{{LIGHT-SABRE}} Enables Efficient in-Magnet Catalytic Hyperpolarization},
  author = {Theis, Thomas and Truong, Milton and Coffey, Aaron M. and Chekmenev, Eduard Y. and Warren, Warren S.},
  year = 2014,
  month = nov,
  journal = {Journal of Magnetic Resonance},
  volume = {248},
  pages = {23--26},
  issn = {1090-7807},
  doi = {10.1016/j.jmr.2014.09.005},
  urldate = {2026-01-26}
}

@article{theisMicroteslaSABREEnables2015,
  title = {Microtesla {{SABRE Enables}} 10\% {{Nitrogen-15 Nuclear Spin Polarization}}},
  author = {Theis, Thomas and Truong, Milton L. and Coffey, Aaron M. and Shchepin, Roman V. and Waddell, Kevin W. and Shi, Fan and Goodson, Boyd M. and Warren, Warren S. and Chekmenev, Eduard Y.},
  year = 2015,
  month = feb,
  journal = {Journal of the American Chemical Society},
  volume = {137},
  number = {4},
  pages = {1404--1407},
  publisher = {American Chemical Society},
  issn = {0002-7863},
  doi = {10.1021/ja512242d},
  urldate = {2026-01-26}
}

@article{tomhonTemperatureCyclingEnables2022,
  title = {Temperature {{Cycling Enables Efficient 13C SABRE-SHEATH Hyperpolarization}} and {{Imaging}} of [1-{{13C}}]-{{Pyruvate}}},
  author = {TomHon, Patrick and Abdulmojeed, Mustapha and Adelabu, Isaiah and Nantogma, Shiraz and Kabir, Mohammad Shah Hafez and Lehmkuhl, S{\"o}ren and Chekmenev, Eduard Y. and Theis, Thomas},
  year = 2022,
  month = jan,
  journal = {Journal of the American Chemical Society},
  volume = {144},
  number = {1},
  pages = {282--287},
  publisher = {American Chemical Society},
  issn = {0002-7863},
  doi = {10.1021/jacs.1c09581},
  urldate = {2026-01-26}
}

@article{tratzmillerParallelSelectiveNuclearspin2021,
  title = {Parallel Selective Nuclear-Spin Addressing for Fast High-Fidelity Quantum Gates},
  author = {Tratzmiller, Benedikt and Haase, Jan F. and Wang, Zhenyu and Plenio, Martin B.},
  year = 2021,
  month = jan,
  journal = {Physical Review A},
  volume = {103},
  number = {1},
  pages = {012607},
  publisher = {American Physical Society},
  doi = {10.1103/PhysRevA.103.012607},
  urldate = {2026-01-26}
}

@article{truong15NHyperpolarizationReversible2015,
  title = {{{15N Hyperpolarization}} by {{Reversible Exchange Using SABRE-SHEATH}}},
  author = {Truong, Milton L. and Theis, Thomas and Coffey, Aaron M. and Shchepin, Roman V. and Waddell, Kevin W. and Shi, Fan and Goodson, Boyd M. and Warren, Warren S. and Chekmenev, Eduard Y.},
  year = 2015,
  month = apr,
  journal = {The Journal of Physical Chemistry C},
  volume = {119},
  number = {16},
  pages = {8786--8797},
  publisher = {American Chemical Society},
  issn = {1932-7447},
  doi = {10.1021/acs.jpcc.5b01799},
  urldate = {2026-01-26}
}

@article{barskiySimpleAnalyticalModel2015,
  title = {A Simple Analytical Model for Signal Amplification by Reversible Exchange ({{SABRE}}) Process},
  author = {Barskiy, Danila A. and Pravdivtsev, Andrey N. and Ivanov, Konstantin L. and Kovtunov, Kirill V. and Koptyug, Igor V.},
  year = 2015,
  month = dec,
  journal = {Physical Chemistry Chemical Physics},
  volume = {18},
  number = {1},
  pages = {89--93},
  publisher = {The Royal Society of Chemistry},
  issn = {1463-9084},
  doi = {10.1039/C5CP05134G},
  urldate = {2025-02-17},
  langid = {english}
}

@article{rodinRepresentationPopulationExchange2020,
  title = {Representation of Population Exchange at Level Anti-Crossings},
  author = {Rodin, B. A. and Ivanov, K. L.},
  year = 2020,
  journal = {Magnetic Resonance},
  volume = {1},
  number = {2},
  pages = {347--365},
  doi = {10.5194/mr-1-347-2020}
}

@article{kiryutinCompleteMagneticField2018,
  title = {Complete Magnetic Field Dependence of {{SABRE-derived}} Polarization},
  author = {Kiryutin, Alexey S. and Yurkovskaya, Alexandra V. and Zimmermann, Herbert and Vieth, Hans-Martin and Ivanov, Konstantin L.},
  date = {2018-07},
  journaltitle = {Magnetic Resonance in Chemistry},
  shortjournal = {Magn Reson Chem},
  volume = {56},
  number = {7},
  pages = {651--662},
  issn = {07491581},
  doi = {10.1002/mrc.4694},
  url = {https://onlinelibrary.wiley.com/doi/10.1002/mrc.4694},
  urldate = {2023-02-08},
  langid = {english}
}

@article{mewisStrategiesHyperpolarizationAcetonitrile2015,
  title = {Strategies for the {{Hyperpolarization}} of {{Acetonitrile}} and {{Related Ligands}} by {{SABRE}}},
  author = {Mewis, Ryan E. and Green, Richard A. and Cockett, Martin C. R. and Cowley, Michael J. and Duckett, Simon B. and Green, Gary G. R. and John, Richard O. and Rayner, Peter J. and Williamson, David C.},
  year = 2015,
  month = jan,
  journal = {The Journal of Physical Chemistry B},
  volume = {119},
  number = {4},
  pages = {1416--1424},
  publisher = {American Chemical Society},
  issn = {1520-6106},
  doi = {10.1021/jp511492q},
  urldate = {2026-02-11}
}

@article{knechtTheoreticalDescriptionHyperpolarization2020,
  title = {Theoretical Description of Hyperpolarization Formation in the {{SABRE-relay}} Method},
  author = {Knecht, Stephan and Barskiy, Danila A. and Buntkowsky, Gerd and Ivanov, Konstantin L.},
  year = 2020,
  month = oct,
  journal = {The Journal of Chemical Physics},
  volume = {153},
  number = {16},
  pages = {164106},
  issn = {0021-9606},
  doi = {10.1063/5.0023308},
  urldate = {2026-02-16}
}
\bibliographystyle{rsc} 

\end{document}


\vspace*{.01 in}
\maketitle
\vspace{.12 in}

\vspace{.12 in}

\section{Experimental SABRE optimisations}
\subsection{SABRE-SHEATH}

\begin{figure}[H]
\centering
  \includegraphics[width=0.45\textwidth]{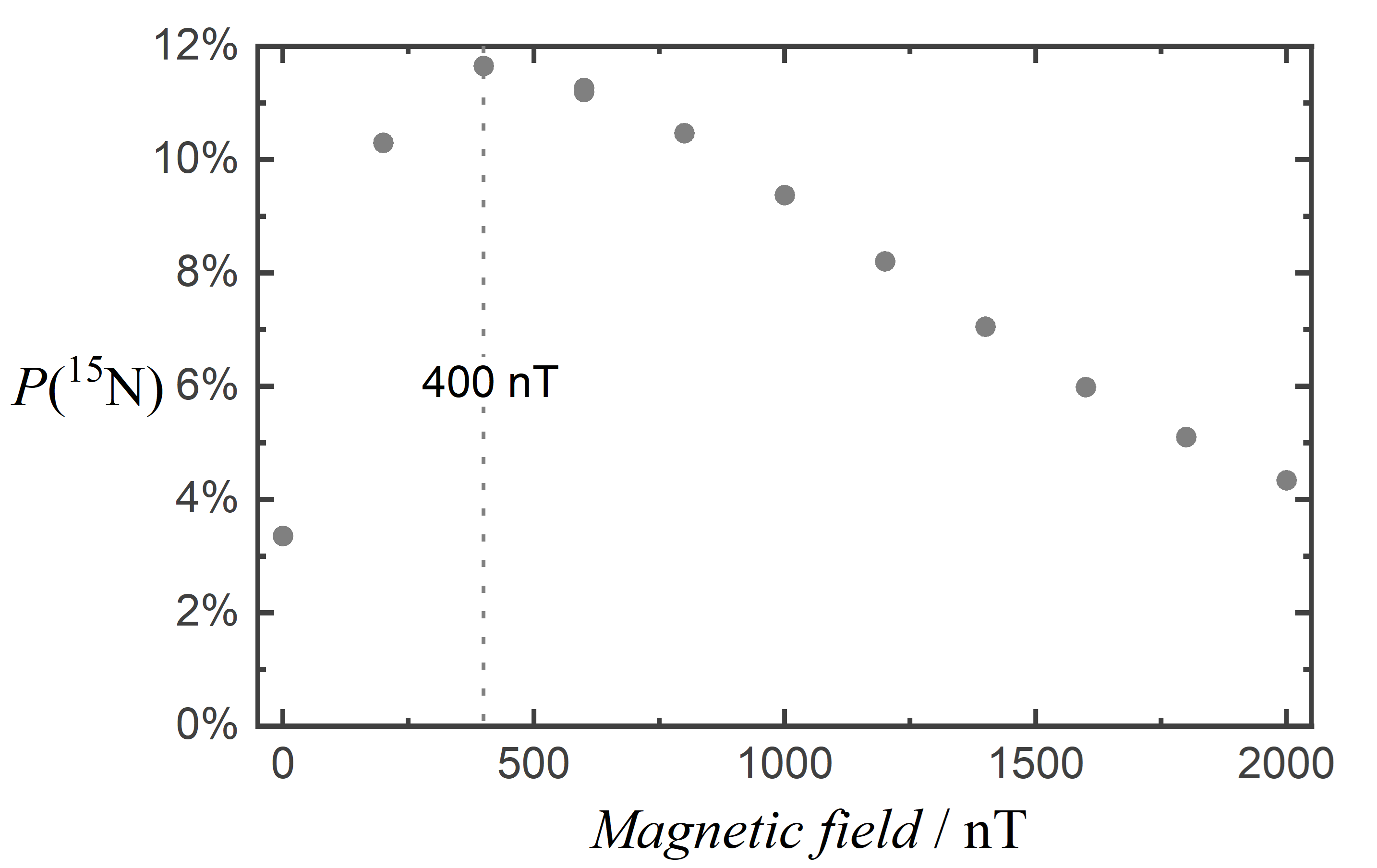}
  \caption{SABRE-SHEATH optimization of \Nlab polarisation achieved on \Nlab-acetonitrile at different bias fields during bubbling with parahydrogen. Experiments were performed at room temperature. Optimal SABRE-SHEATH field is indicated.}
  \label{fgr:SI_sheath_opt} 
\end{figure}

SABRE-SHEATH was optimised by varying the static magnetic fields at which SABRE polarisation takes places. The example data is shown in Fig.~\ref{fgr:SI_sheath_opt} and the optimised values are given in the table below.

~

 \begin{table}[h]
 \caption{Optimal \textbf{SABRE-SHEATH} bias fields and \Nlab polarisation levels achieved at room temperature. Errors were estimated by performing experiments six times.}
 \centering
 \begin{tabular}{||c||c c||} 
 \hline
 System & Optimal field & \Nlab polarisation \\ [0.5ex] 
 \hline\hline
 \Nlab-acetonitrile  & 400~nT & $11.7\pm0.4\%$ \\ 
 \hline
 \Nlab-pyridine & 400~nT & $6.8\pm0.2\%$ \\ 
 \hline
 Metronidazole  & 800~nT & $9.1\pm0.6\%$ \\ 
 \hline
\end{tabular}\label{table:SABRE-SHEATH}
\end{table}

\subsection{SABRE-SLIC}

\begin{figure}[H]
\centering
  \includegraphics[width=0.45\textwidth]{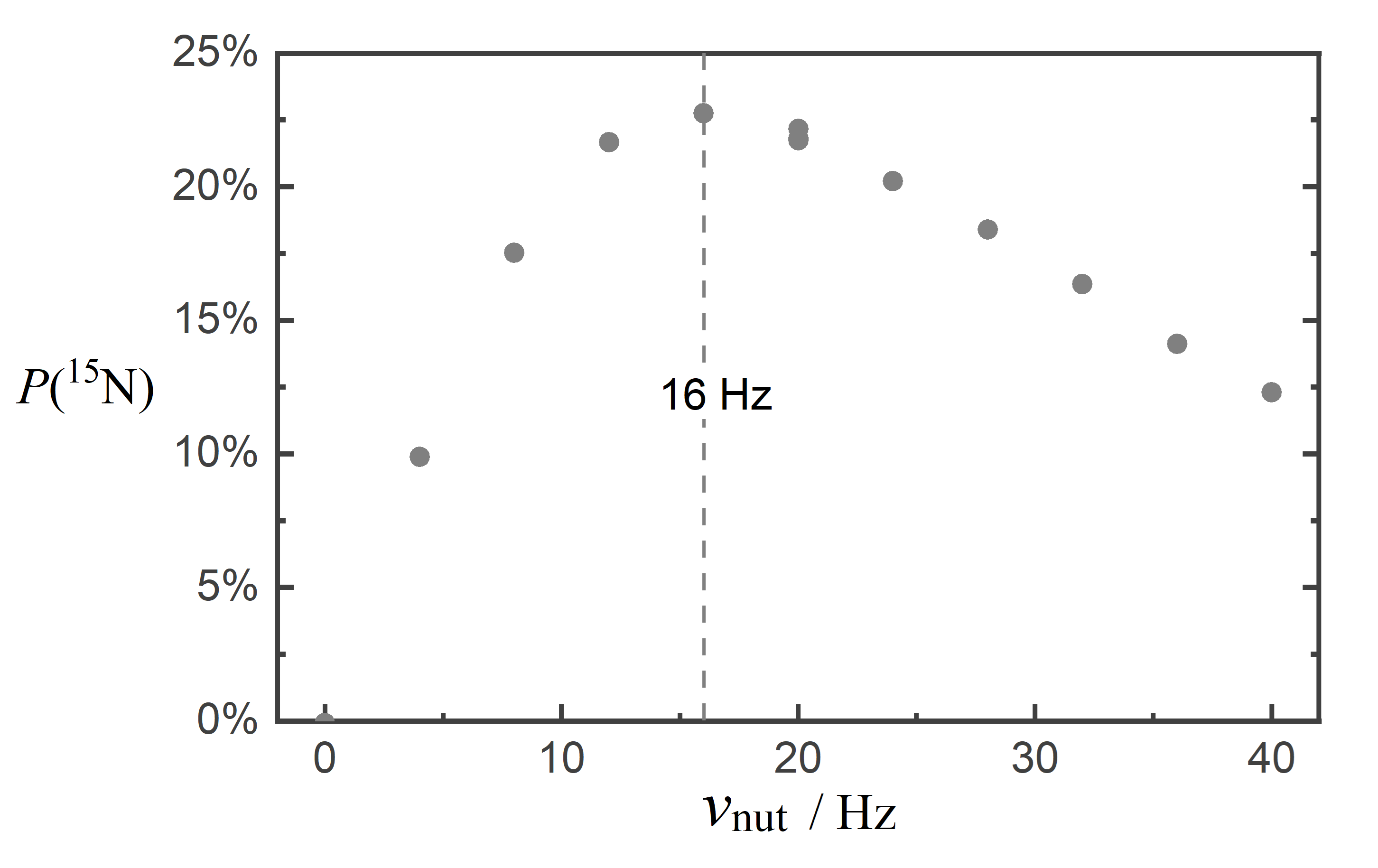}
  \caption{SABRE-SLIC optimization of \Nlab polarisation achieved on \Nlab-acetonitrile at different oscillating field $B_1$ amplitudes with bias field set to 98~$\mu$T. Experiments were performed at room temperature. Optimal amplitude is indicated.}
  \label{fgr:SI_SLIC_opt} 
\end{figure}

SABRE-SLIC was optimised by varying the amplitude of the oscillating magnetic field at a bias field of 98~$\mu$T. The frequency was set to the \Nlab resonance frequency. The example data is shown in Fig.~\ref{fgr:SI_SLIC_opt} and the optimised values are given in the table below.

~

 \begin{table}[h]
 \caption{Optimal \textbf{SABRE-SLIC} $B_1$ amplitudes and \Nlab polarisation levels achieved at room temperature. Experiments were performed at fixed 98~$\mu$T bias field. Errors were estimated by repeating experiments four times.}
 \centering
 \begin{tabular}{||c||c c||} 
 \hline
 System & $B_1$ amplitudes & \Nlab polarisation \\ [0.5ex] 
 \hline\hline
 \Nlab-acetonitrile  & 16~Hz & $21.1\pm0.3\%$ \\ 
 \hline
 \Nlab-pyridine & 16~Hz & $15.2\pm0.9\%$ \\ 
 \hline
 Metronidazole  & 16~Hz & $12.9\pm0.2\%$ \\ 
 \hline
\end{tabular}\label{table:SABRE-SLIC}
\end{table}

\subsection{SABRE-DRF-SLIC}

\begin{figure}[H]
\centering
  \includegraphics[width=0.45\textwidth]{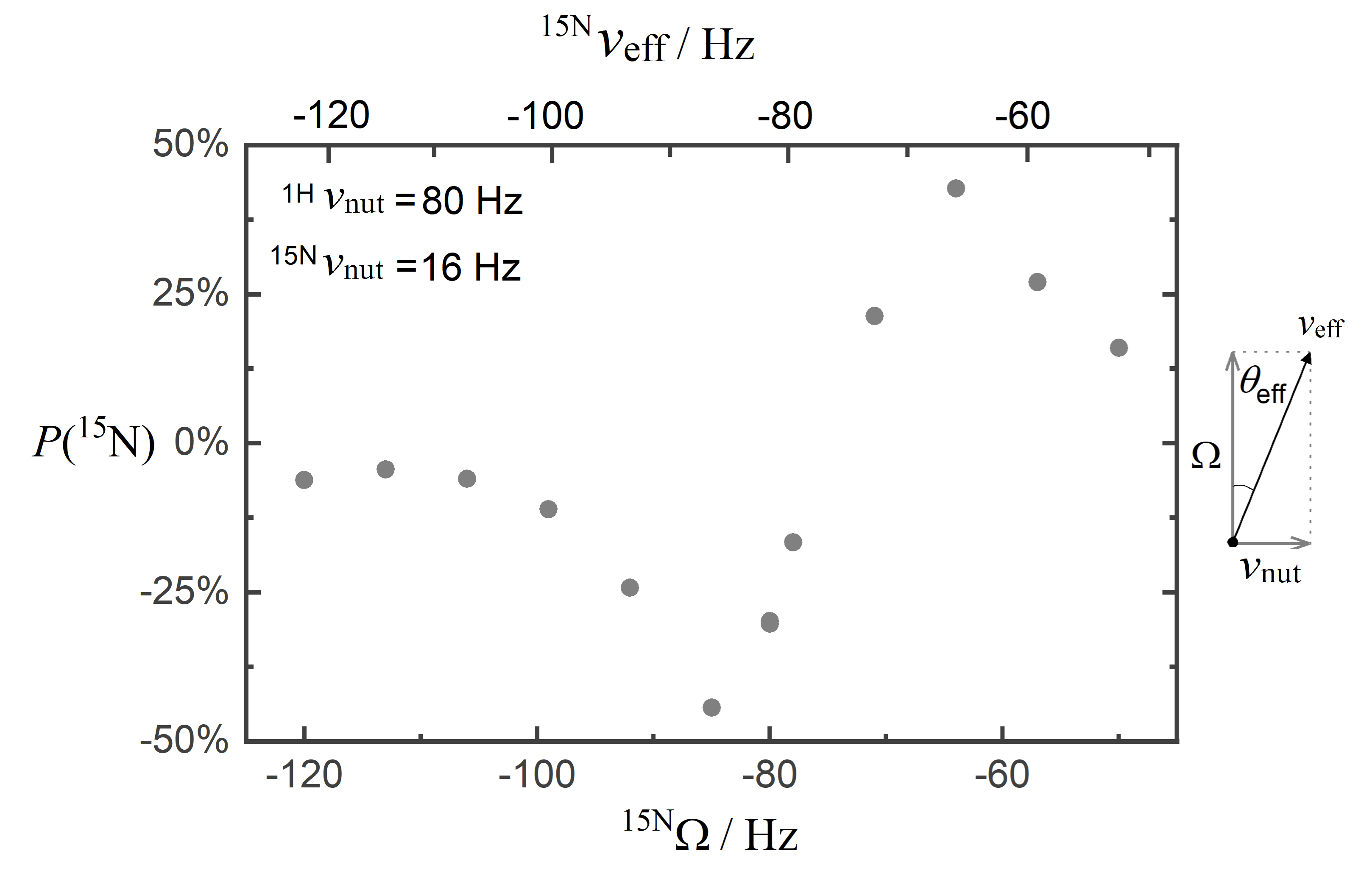}
  \caption{SABRE-DRF-SLIC optimization of \Nlab polarisation achieved on \Nlab-acetonitrile at different oscillating field frequency offsets. The bias magnetic field was set to 98~$\mu$T. Experiments were performed at room temperature. The effective angle is depicted on the right for clarity.}
  \label{fgr:SI_DRFSLIC_opt} 
\end{figure}

SABRE-DRF-SLIC experiments were performed by utilizing two transverse oscillating fields. The first was set to be resonant with \Hone at amplitude of 80~Hz. The second field was set to amplitude of 16~Hz and with a slight mismatch with the \Nlab Larmor frequency. The mismatch was varied and performance profiles were recorded as in the Fig.~\ref{fgr:SI_DRFSLIC_opt}. The optimal values together which the effective angle of the effective field for \Nlab spins are indicated in the table below.

~

 \begin{table}[H]
 \caption{Optimal \textbf{SABRE-DRF-SLIC} parameters and \Nlab polarisation levels achieved at room temperature. Experiments were performed at fixed 96~$\mu$T bias field. Errors were estimated by repeating experiments four times. Number of cycles was adjusted for overall duration of 30~s.}
 \centering
 \begin{tabular}{||l||c c c c c||} 
 \hline
 System & $^{1H}\nu_\mathrm{nut}$  & $^{15N}\nu_\mathrm{nut}$  & $^{15N}\Omega$ & $\theta_{\mathrm{eff}}$ & \Nlab polarisation \\ [0.5ex] 
 \hline\hline
 \Nlab-acetonitrile  & 80~Hz & 16~Hz & $-64$~Hz & 14.0\degr &$45.3\pm0.3\%$ \\ 
 \hline
 \Nlab-pyridine & 80~Hz & 16~Hz & $-64$~Hz & 14.0\degr &$23.0\pm1.6\%$ \\ 
 \hline
 Metronidazole & 80~Hz & 30~Hz & $-104$~Hz & 16.1\degr &$3.7\pm0.1\%$ \\ 
 \hline
\end{tabular}\label{table:SABRE-DRF-SLIC}
\end{table}

\subsection{SABRE-PulsePol}

\begin{figure}[H]
\centering
  \includegraphics[width=0.45\textwidth]{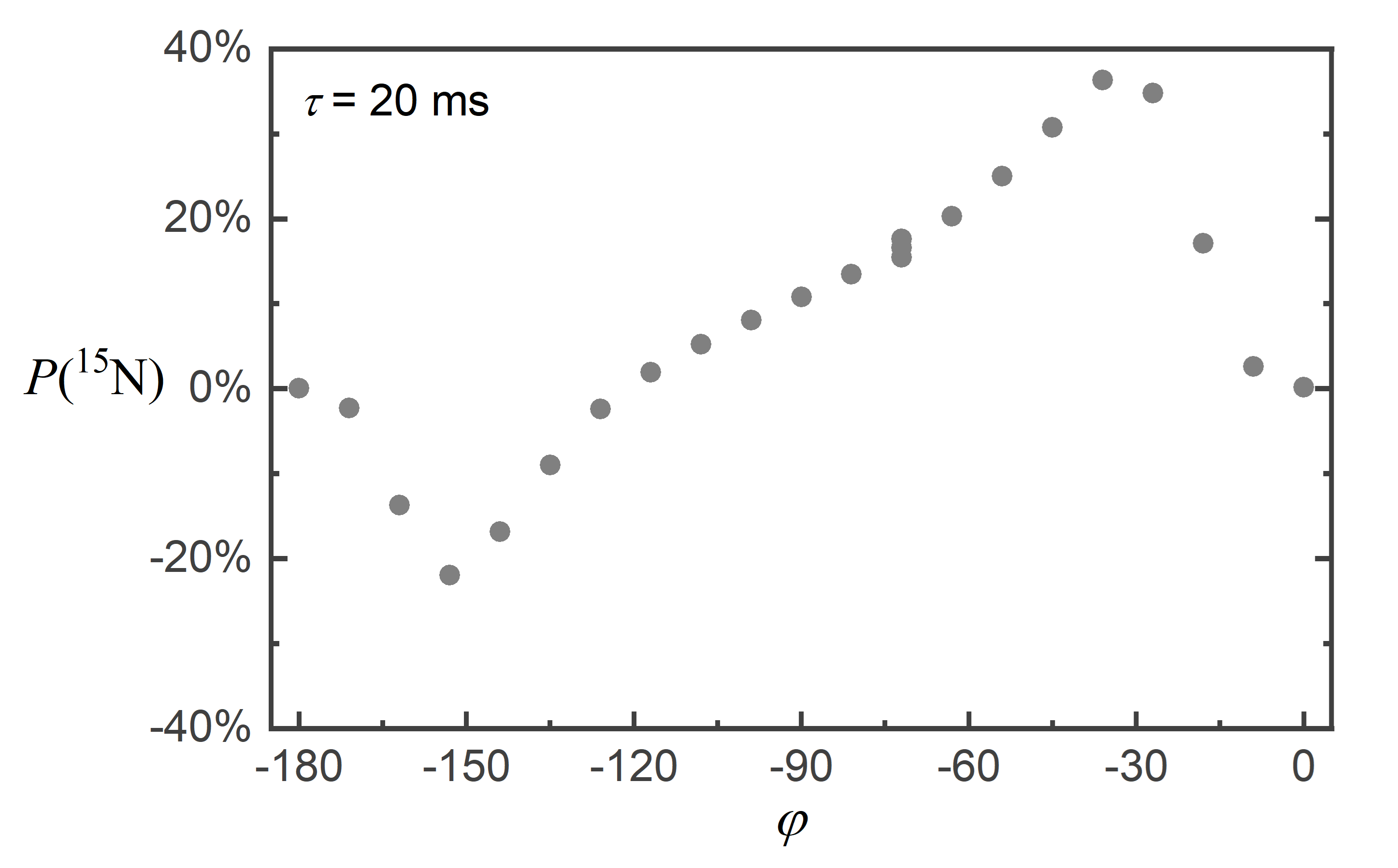}
  \caption{SABRE-PulsePol optimization of \Nlab polarisation achieved on \Nlab-acetonitrile at different phase $\varphi$ settings at fixed PulsePol cycle time $\tau$. The bias magnetic field was set to 1000~$\mu$T. }
  \label{fgr:SI_PP_opt} 
\end{figure}

SABRE-PulsPol experiments were performed by utilizing a single channel sequence. The pulses were resonant with \Nlab spins with amplitude of 600~Hz. The bias field was elevated to 1~mT to avoid Bloch-Siegert shift. The pulse shape was set to Gaussian with 10\% truncation to avoid possible \Hone excitation. Experimental optimization involved exploration of two parameter space, the phase $\varphi$ and evolution time $\tau$. The sequence is the depicted in the main text. An profile acquired by varying phase at fixed evolution time is shown in Fig.~\ref{fgr:SI_PP_opt} and the optimized values are provided in the table below.

~

 \begin{table}[h]
 \caption{Optimal \textbf{SABRE-PulsePol} phase step and cycle times and \Nlab polarisation levels achieved at room temperature. Experiments were performed at fixed 1~mT bias field. Errors were estimated by repeating experiments four times. Number of cycles was adjusted for overall duration of 30~s.}
 \centering
 \begin{tabular}{||c||c c c||} 
 \hline
 System & Phase step $\varphi$ & Cycle time $\tau_\mathrm{PP}$ & \Nlab polarisation \\ [0.5ex] 
 \hline\hline
 \Nlab-acetonitrile  & -36\degr & 20.0~ms & $48.6\pm0.3\%$ \\ 
 \hline
 \Nlab-pyridine & -36\degr & 22.2~ms & $29.4\pm1.6\%$ \\  
 \hline
 Metronidazole  & -72\degr & 16.0~ms & $4.5\pm0.5\%$ \\ 
 \hline
\end{tabular}\label{table:SABRE-PulsePol}
\end{table}

\newpage

\section*{SABRE at lower temperature}

\begin{figure}[h]
\centering
  \includegraphics[width=0.45\textwidth]{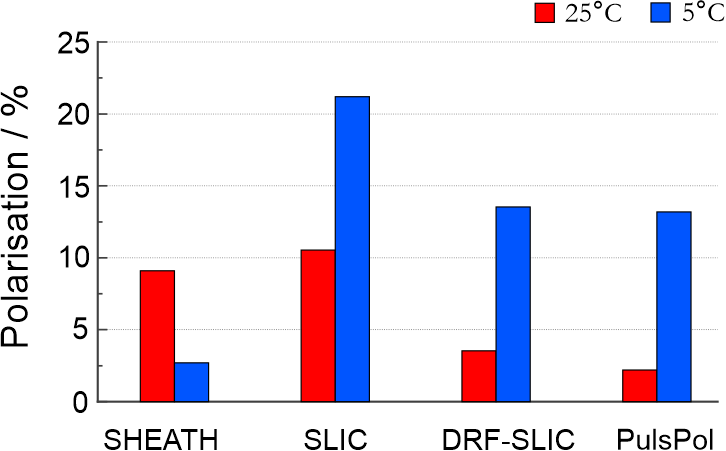}
  \caption{\Nlab polarisation levels achieved on MTZ using different SABRE methods at 25\degC and 5\degC.}
  \label{fgr:SI_tempMTZ} 
\end{figure}

We have also ran an additional experiments on metronidazole system at reduced temperature of 5\degC and optimized each SABRE method as previously described. The results are summarized in Fig.~\ref{fgr:SI_tempMTZ}. We have observed an overall increase in PulsPol and DRF-SLIC performance at lower temperature. However, the maximum \Nlab polarisation was still acquired using SABRE-SLIC which suggests that the exchange rate may be expected to be relatively high which experimentally was determined to be around 60 1/s.

\section{Numerical Calculations}

To perform the numerical simulations, the superoperator approach was used. It allows for the accounting of coherent, relaxation, and exchange dynamics. The exact theoretical details can be found elsewhere~\cite{knechtTheoreticalDescriptionHyperpolarization2020}, and the exact implementation can be found in the repository link - \note{[repository link to be added]}. For the calculation simplicity, the rapid parahydrogen exchange is assumed.  In this analysis, we considered a single $^{15}$N nucleus in the substrate, while the SABRE complex yields a 3-spin system with two additional protons derived from parahydrogen. For the case of continuous wave methods, the steady-state solution was derived from the null-space. For pulse methods (PulsePol), 50 seconds of evolution time was granted. The $J_\mathrm{HH}=-7.7$ Hz and the $J_\mathrm{NH}=-21$ Hz. Relaxation was phenomenological, with $T_1$ = 6 seconds on protons and 20 seconds on nitrogen.

\subsection{SHEATH}

\begin{figure}[H]
\centering
  \includegraphics[width=0.45\textwidth]{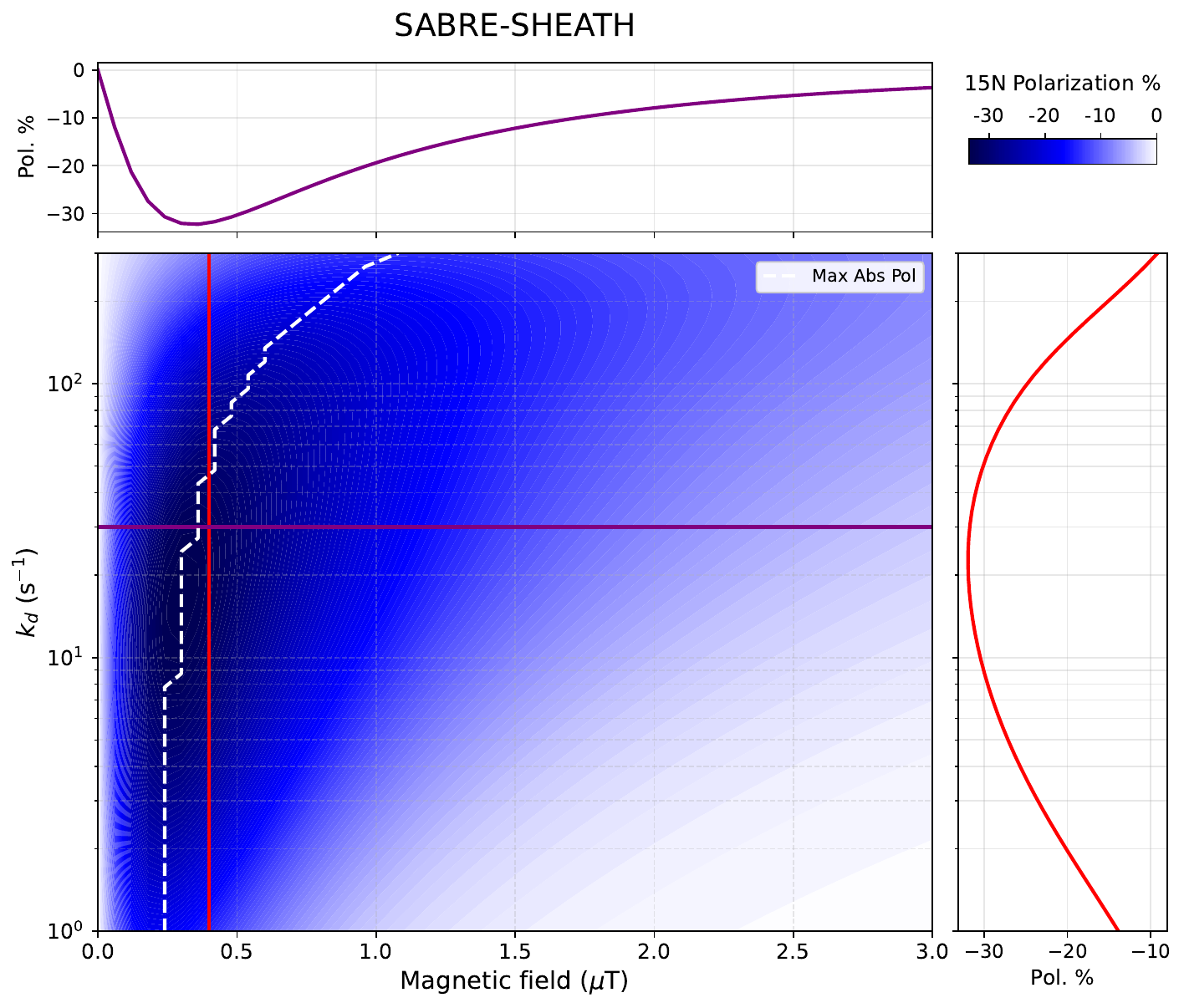}
  \caption{
Simulated $^{15}$N SABRE-SHEATH polarisation efficiency. The main central panel displays a 2D contour map of the final $^{15}$N polarisation as a function of the external magnetic field ($\mu$T) and the chemical exchange rate $k_d$ (s$^{-1}$, presented on a logarithmic scale). The \textit{white dashed line} traces the magnetic field value yielding the maximum absolute polarisation for each exchange rate. The color scale indicates negative polarisation, with darker blue regions representing higher absolute transfer efficiencies (approaching -30\%). The top panel presents a 1D horizontal cross-section illustrating the magnetic field dependence at a fixed optimal exchange rate of $k_d = 30$ s$^{-1}$, corresponding to the solid purple line in the 2D map. The right panel presents a 1D vertical cross-section detailing the exchange rate dependence at a fixed matching field of 0.4 $\mu$T, denoted by the solid red line. 
}
  \label{fgr:SABRE_SHEATH} 
\end{figure}

The simulation results for SABRE-SHEATH are shown in Figure~\ref{fgr:SABRE_SHEATH}. In the slow-to-intermediate exchange regime, the optimal $B_0$ values lie between 200 and 600 nT. This coincides well with our experimental data, see Table~\ref{table:SABRE-SHEATH}, where all investigated compounds exhibited an optimal amplitude of 400 nT, with the exception of 800 nT for Metronidazole which has the largest $k_\mathrm{d}$.  Interestingly, the optimal $B_0$ experiences a sharp increase beyond $k_d \approx 90$ s$^{-1}$. In this fast-exchange region, the system deviates from the standard Level Anti-Crossing (LAC) condition, effectively sacrificing maximum polarisation amplitude in favor of faster transfer rates to compete with the rapid exchange.

\subsection{SLIC}

\begin{figure}[H]
\centering
  \includegraphics[width=0.45\textwidth]{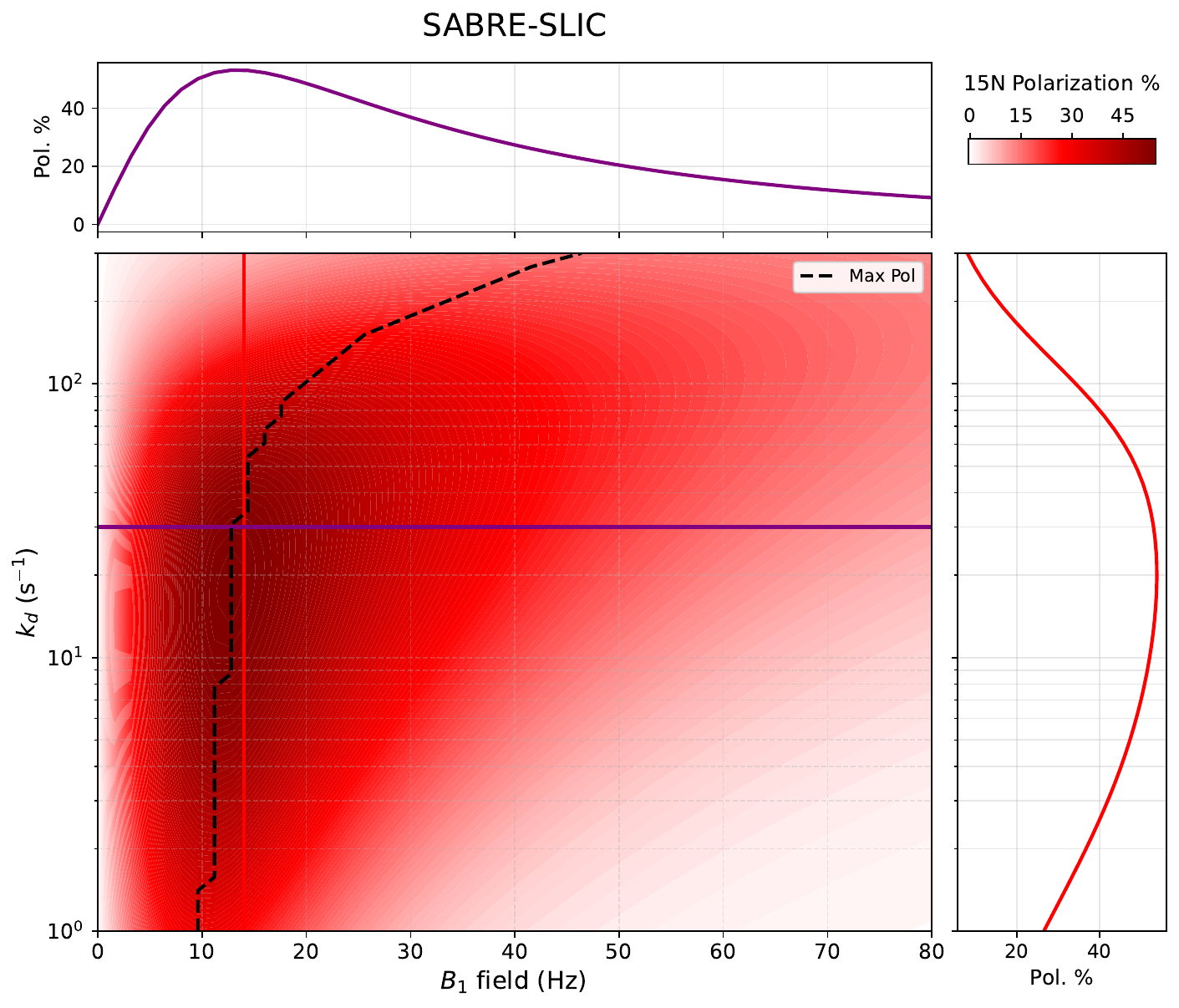}
  \caption{Simulated $^{15}$N SABRE-SLIC polarisation efficiency. The main central panel displays a 2D contour map of the final $^{15}$N polarisation percentage as a function of the applied continuous-wave RF field amplitude ($B_1$, Hz) and the chemical exchange rate $k_d$ (s$^{-1}$, presented on a logarithmic scale). The color scale indicates positive polarisation, with darker red regions representing higher transfer efficiencies (approaching 50\%). The black dashed line traces the optimal RF amplitude ($B_1$) required to maximize polarisation for each specific exchange rate. The top panel presents a 1D horizontal cross-section illustrating the $B_1$ field dependence at a fixed optimal exchange rate of $k_d \approx 30$ s$^{-1}$, corresponding to the solid purple line in the 2D map. The right panel presents a 1D vertical cross-section detailing the exchange rate dependence at a fixed matching RF field of $B_1 \approx 21$ Hz, denoted by the solid red line.}
  \label{fgr:SLIC_NULLSPACE_plot} 
\end{figure}

The simulation results for SABRE-SLIC are shown in Figure~\ref{fgr:SLIC_NULLSPACE_plot}. In the slow-to-intermediate exchange regime, the optimal $B_1$ values lie between 10 and 20 Hz. This coincides well with our experimental data, see Table~\ref{table:SABRE-SLIC}, where all investigated compounds exhibited an optimal amplitude near 16 Hz, despite the expectation that Metronidazole (with the largest $k_\mathrm{d}$) might require a higher value. Interestingly, the optimal $B_1$ experiences a sharp increase beyond $k_d \approx 90$ s$^{-1}$. The reason it is happening is exactly identical to the case of $B_0$ variation in the SABRE-SHEATH.

\begin{figure}[H]
\centering
  \includegraphics[width=0.9\textwidth]{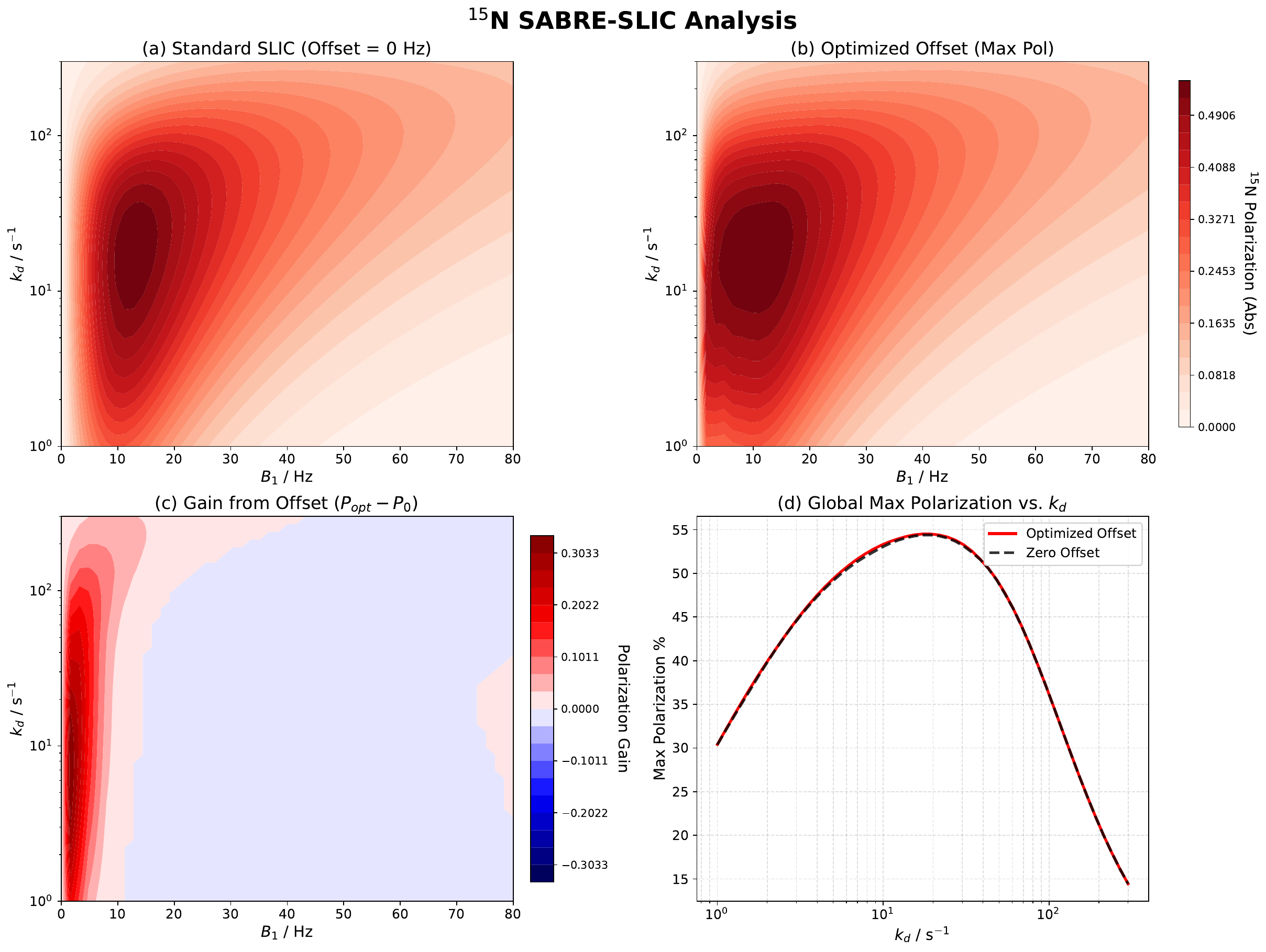}
  \caption{Impact of RF offset frequency on SABRE-SLIC performance.
(a) Simulated $^{15}$N polarisation map for standard SLIC (zero offset).
(b) Polarization map obtained by optimizing the RF offset frequency at each $(B_1, k_d)$ coordinate.
(c) The difference map ($|P_{\text{opt}}| - |P_{\text{zero}}|$), showing negligible gains across the relevant parameter space.
(d) Maximum achievable polarisation as a function of the exchange rate $k_d$. The close overlap between the optimized (solid red) and zero-offset (dashed black) curves confirms that adding an offset does not significantly improve transfer efficiency in the SLIC regime.}
  \label{fgr:SLIC_Offset_Comparison_2x2} 
\end{figure}

The introduction of an RF offset does not yield a global increase in polarisation, as demonstrated in Figure~\ref{fgr:SLIC_Offset_Comparison_2x2}. An enhancement in transfer efficiency is observed only in the low $B_1$ regime, likely due to the realignment of the Level Anti-Crossing (LAC) condition. A detailed theoretical analysis of off-resonant SLIC is provided in the main text, which is consistent with previously reported results~\cite{kozinenkoSLICSABREMicroteslaFields2025}. To address the limitations of the standard sequence, we propose the DRF-SLIC pulse sequence, which is detailed in the following section.

\subsection{DRF-SLIC}

\begin{figure}[H]
\centering
  \includegraphics[width=0.9\textwidth]{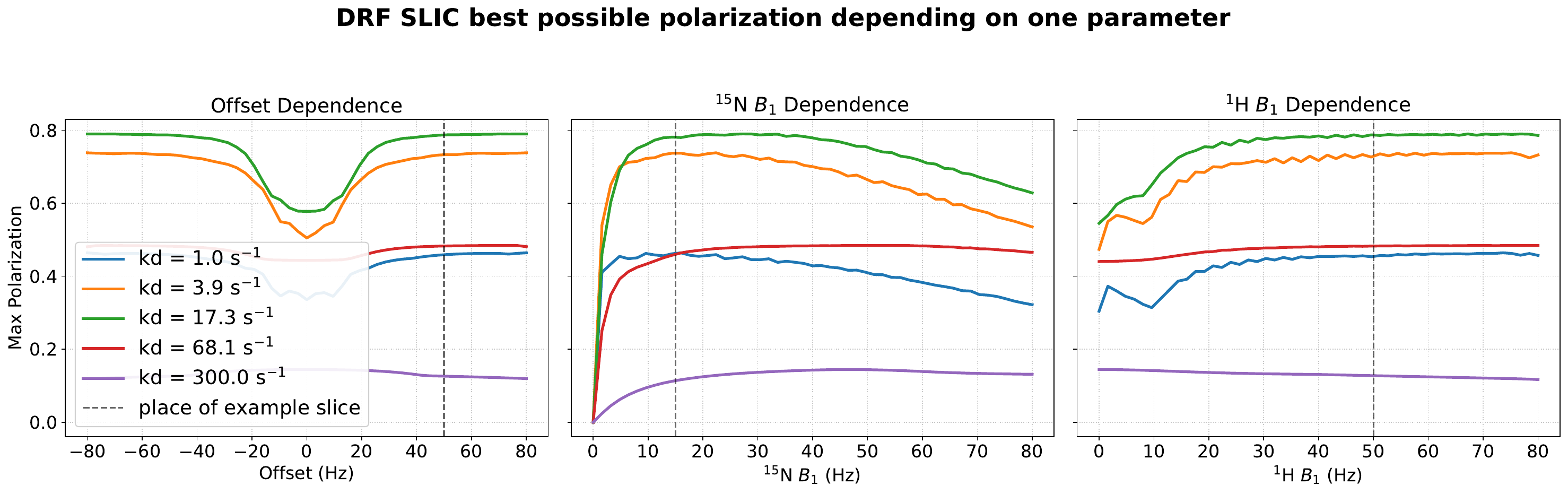}
  \caption{The panels display the maximum simulated $^{15}$N polarisation as a function of a single control variable. The dependence is shown for \textit{left} the resonance offset $\Delta_N$, \textit{center} the nitrogen RF amplitude $B_{1N}$, and \textit{right} the proton RF amplitude $B_{1H}$. The colored curves represent different chemical exchange rates ($k_d$) ranging from slow ($1.0$ s$^{-1}$) to fast ($300.0$ s$^{-1}$) exchange. Vertical dashed lines indicate the parameter set ($\Delta_N=50$ Hz, $B_{1N}=15$ Hz, $B_{1H}=50$ Hz) selected for 2D analysis. The data indicate that while $B_{1N}$ requires precise calibration (10–20 Hz), the $\Delta_N$ and $B_{1H}$ parameters exhibit broad tolerance, simplifying the experimental search space.}
  \label{fgr:DRF_SLIC_best_pol_one_parameter} 
\end{figure}

Since the DRF-SLIC method relies on three independent control parameters, experimentally mapping the entire parameter space would be time-consuming. However, the optimization scans in Figure~\ref{fgr:DRF_SLIC_best_pol_one_parameter} reveal a convenient optimisation strategy. The optimal nitrogen nutation field ($B_\mathrm{1N}$) remains confined to a narrow, predictable window between 10 and 20 Hz, regardless of the exchange rate. Conversely, the nitrogen offset ($\Delta_\mathrm{N}$) and proton nutation field ($B_\mathrm{1H}$) show a broad efficiency plateau; any values exceeding 30 Hz yield an increase in polarisation, with efficiency growing slowly but steadily as the fields increase. This suggests that experimental optimization can be simplified by fixing $B_\mathrm{1N}$ near 15 Hz and focusing solely on maximizing the remaining two parameters.

\begin{figure}[H]
\centering
  \includegraphics[width=0.9\textwidth]{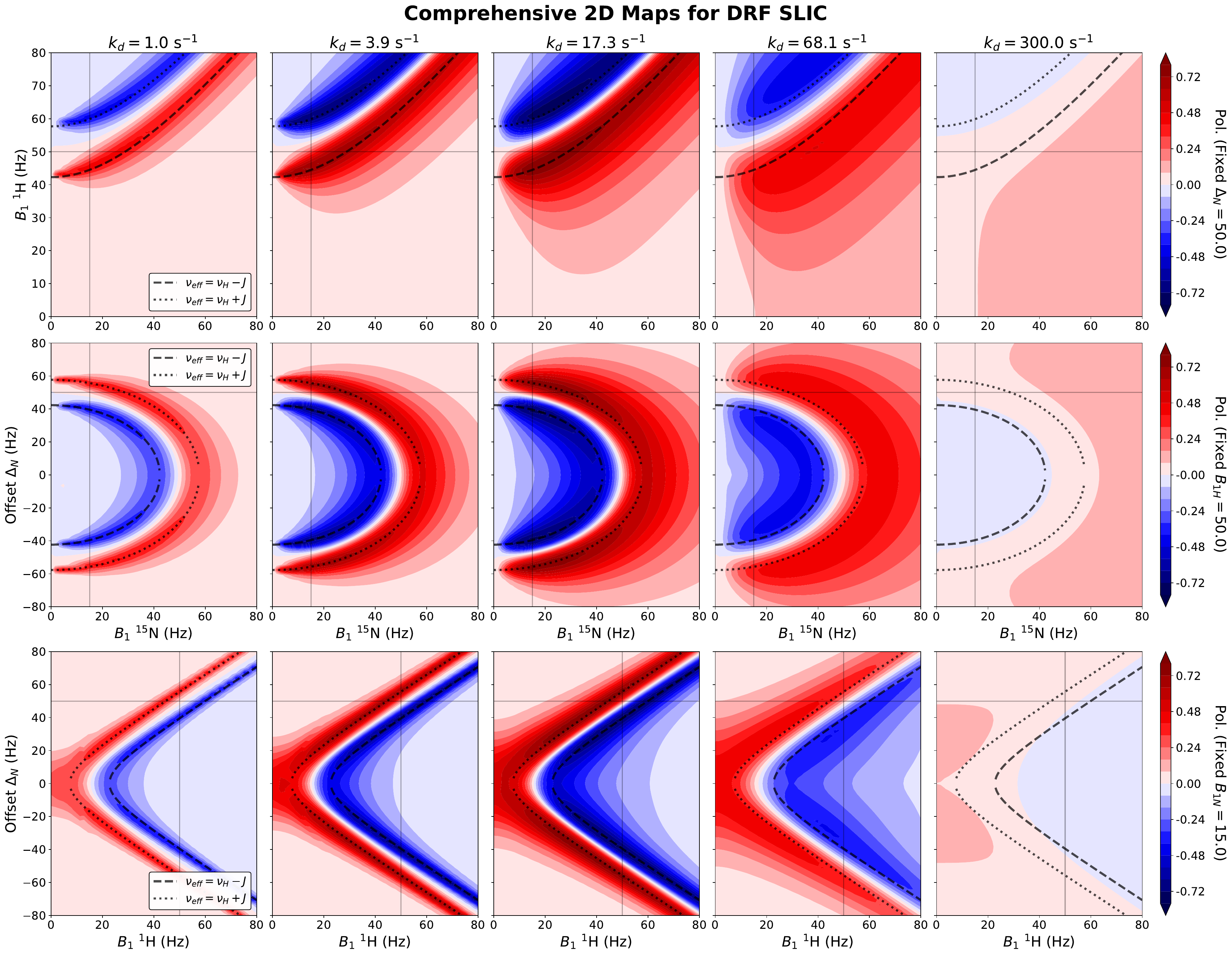}
  \caption{Simulated 2D maps of $^{15}$N polarisation transfer efficiency under DRF-SLIC conditions. The contour plots display the final $^{15}$N polarisation as a function of RF field amplitudes ($B_{1H}$, $B_{1N}$) and resonance offset ($\Delta_N$). The rows present different 2D cross-sections through the parameter space by fixing one variable: (Top Row) fixed offset $\Delta_N = 50$ Hz; (Middle Row) fixed proton RF field $B_{1H} = 50$ Hz; (Bottom Row) fixed nitrogen RF field $B_{1N} = 15$ Hz. The columns correspond to increasing chemical exchange rates $k_d$, from 1.0 s$^{-1}$ (left) to 300.0 s$^{-1}$ (right). The color scale indicates signed polarisation, where red is positive and blue is negative. The overlaid black dashed and dotted curves represent the theoretical Level Anti-Crossing (LAC) conditions ($\nu_{\text{eff}}^N = \nu_{\text{nut}}^H \pm J_{HH}$), predicting the regions of maximal transfer. Thin gray solid lines indicate the fixed parameter values used in the complementary rows. The plots highlight the broadening of the efficient transfer condition as the exchange rate increases.}
  \label{fgr:DRF_SLIC_NULLSPACE_Comprehensive_2D} 
\end{figure}

It seems like it is generally easy to fix two of the parameters from the good value area, take the value for the second one based on the LAC conditions, and then vary leftover on to get the final good polarisation. The comprehensive 2D heatmaps with one parameter fixed and two others being varied is shown in Figure~\ref{fgr:DRF_SLIC_NULLSPACE_Comprehensive_2D}. It is worth noting that the maximal polarisation along LAC lines (dashed line on the figure) is held really well for small values of $k_d$, but become not so strong at larger $k_d$ values, where the requirements for faster evolution may become more important than full population transfer.

The experimental results (see Table~\ref{table:SABRE-DRF-SLIC}) are consistent with the simulations. For a fixed $B_\mathrm{1H}$, the optimal $B_\mathrm{1N}$ lies in the range of 10–20 Hz for compounds with intermediate exchange rates (experimentally 16 Hz for acetonitrile and pyridine) and increases for compounds with faster exchange rates (30 Hz for metronidazole).Furthermore, the optimal offset frequency generally follows the Level Anti-Crossing (LAC) condition for the intermediate exchange regime, where $\nu_\mathrm{eff}^\mathrm{N} \approx \nu_1^\mathrm{H} \pm J_\mathrm{HH}$. For acetonitrile and pyridine, the experimental offset of 64 Hz aligns well with the theoretical prediction of 70 Hz. In contrast, for metronidazole, the experimental offset of 104~Hz deviates from the theoretical LAC value of 82~Hz, likely because the fast exchange dynamics disrupt the standard LAC condition.

\subsection{PulsePol}

The resonance condition for the PulsePol sequence is defined as:

\begin{equation}
    \tau_{\text{PP}} = \frac{2N \pm \varphi /\pi}{J}
\end{equation}

Where $n$ is the resonance number, $J$ corresponds to the scalar coupling $J_\mathrm{HH'}$, and $\varphi$ is the phase shift of the second PulsePol block in radians. The effective coupling strength ($dJ_*$) that drives the polarisation evolution scales by a factor $\alpha$, which depends on the pulse parameters:

\begin{equation}
    dJ_* = dJ \cdot \alpha, \quad \text{where} \quad \alpha = \frac{\sin^2\left( (2n~2\pi \pm 2\varphi) / 8 \right)}{(2n~2\pi \pm 2\varphi)/ 8} \;\text{or} \; \frac{\sin^2{(\pi J\tau_{\text{PP}}/4})}{\pi J\tau_{\text{PP}} / 4}
\end{equation}

 theoretically, the most efficient coupling corresponds to the $n=1_-$ resonance condition, which yields a scaling factor of $\alpha_{1-}(\varphi=\pi/2) \approx 0.725$. However, in our SABRE system, this condition results in a PulsePol cycle duration that is too long relative to the ligand exchange lifetime, effectively damping the coherent transfer dynamics. Consequently, we utilize the faster $n=0$ resonance, despite its lower theoretical scaling factor ($\alpha_{0+}(\varphi=\pi/2) \approx 0.373$).

\begin{figure}[H]
\centering
  \includegraphics[width=0.9\textwidth]{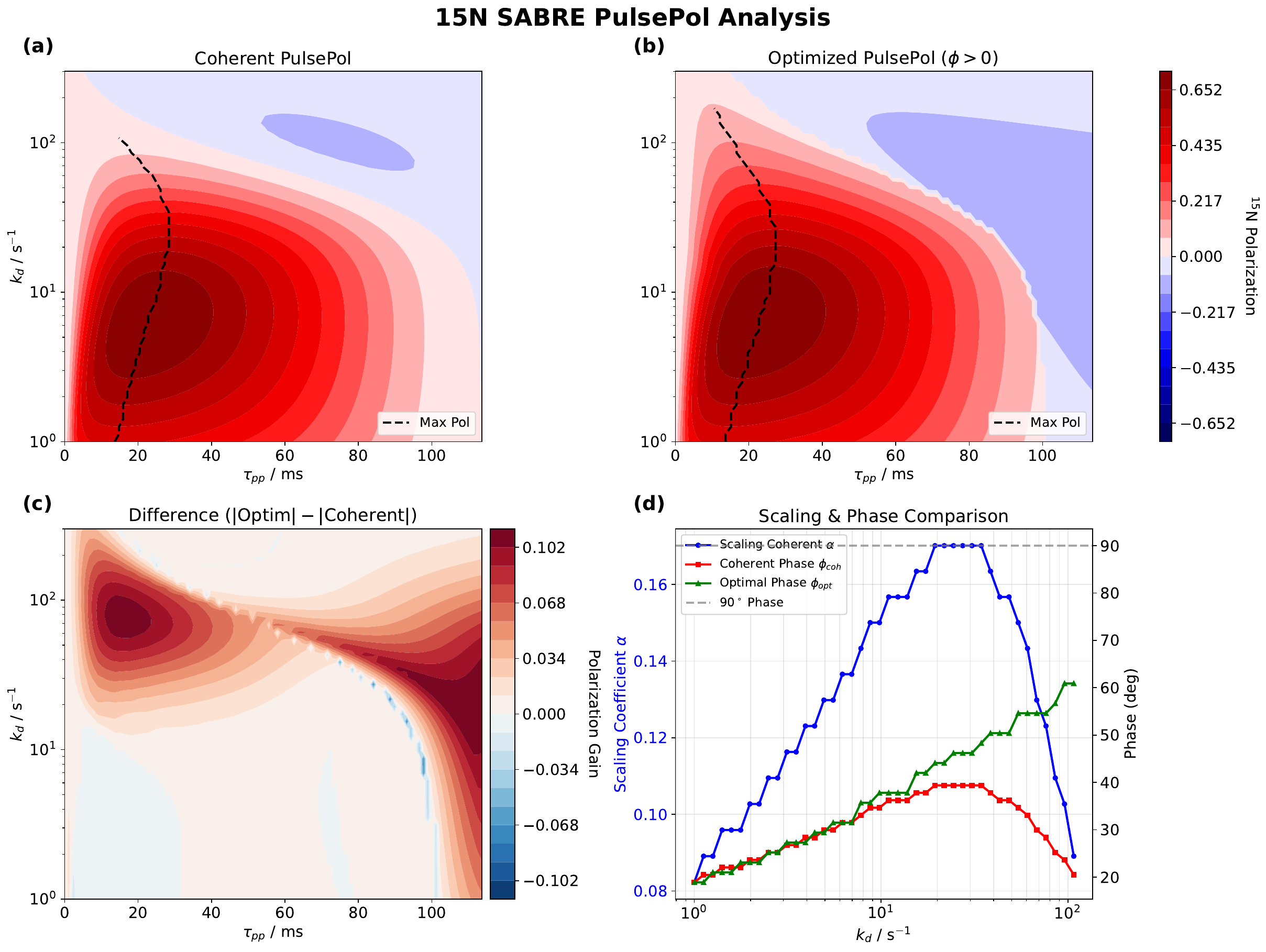}
  \caption{
 Comprehensive analysis of $^{15}$N SABRE PulsePol performance and resonance characteristics.(a) Coherent PulsePol: Simulated $^{15}$N polarisation as a function of the pulse block duration $\tau_{pp}$ and ligand exchange rate $k_d$. The pulse phase $\phi$ is strictly coupled to $\tau_{pp}$ via the theoretical resonance condition $\phi = \pi J \tau_{pp}$ (for the $n=0$ resonance condition). The \textit{black dashed line} traces the trajectory of maximum polarisation ($\tau_{opt}$) for each exchange rate.(b) Optimized PulsePol: The maximum polarisation achievable by numerically optimizing the phase $\phi$ (constrained to $\phi > 0$) at each $(\tau_{pp}, k_d)$ coordinate. This represents the theoretical performance limit when the phase is tuned freely.(c) Polarization Gain: The difference in absolute polarisation magnitude ($|P_{\text{opt}}| - |P_{\text{coh}}|$). Red regions indicate regimes where numerical optimization yields a higher signal than the standard theoretical condition, typically compensating for exchange-induced effects.(d) Scaling and Phase Analysis: A dual-axis comparison of parameters extracted along the maximum polarisation trajectory (dashed lines in a and b).Left Axis (Blue): The theoretical scaling coefficient $\alpha = \sin^2(\pi J \tau_{opt}/4) / (\pi J \tau_{opt}/4)$, calculated using the phase from the coherent condition. This quantifies the theoretical amplitude penalty as the optimal pulse duration increases. Right Axis: A comparison between the theoretical Coherent Phase $\phi_{coh}$ (red squares) and the numerically Optimal Phase $\phi_{opt}$ (green triangles). The divergence of $\phi_{opt}$ from $\phi_{coh}$ at higher $k_d$ illustrates where the system drifts from the ideal resonance condition. The dashed gray line marks $\phi = 90^\circ$.
  }
  \label{fgr:PP_coherent_vs_optim} 
\end{figure}

As shown in Figure~\ref{fgr:PP_coherent_vs_optim}(a), the optimal transfer time $\tau_{\text{opt}}$ initially increases with the exchange rate $k_d$, consistent with the optimal matching condition $k_d \approx 2 \pi dJ_*$. However, at higher exchange rates, the optimal duration begins to decrease. This likely occurs because the exchange lifetime $1/k_d$ becomes comparable to the PulsePol cycle length, forcing the system toward shorter cycles to preserve coherence. 

\begin{figure}[H]
\centering
  \includegraphics[width=0.9\textwidth]{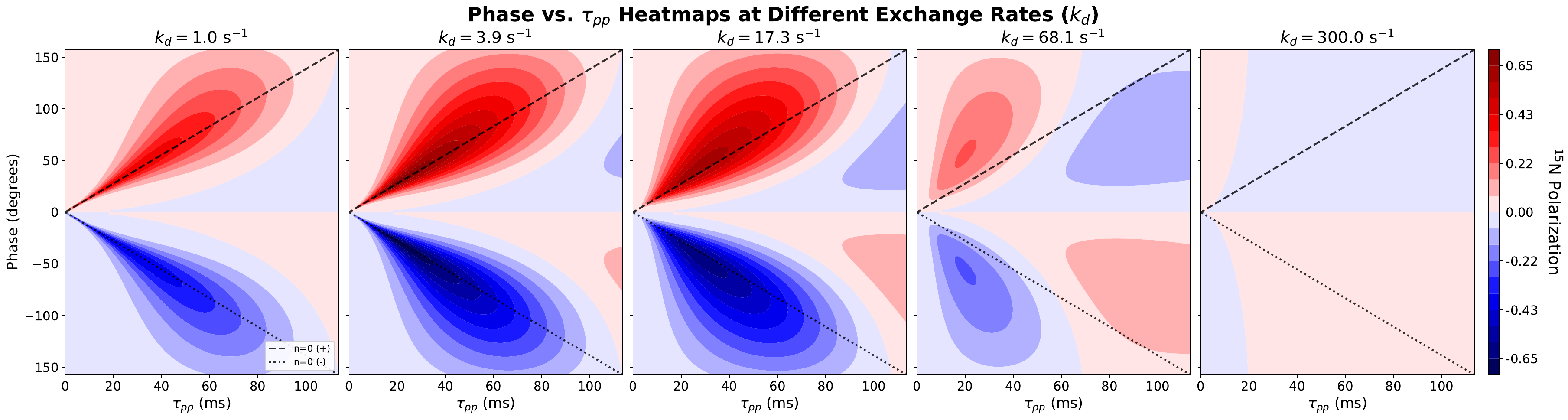}
  \caption{Phase dependence of $^{15}$N polarisation at varying ligand exchange rates.
Simulated $^{15}$N polarisation heatmaps as a function of pulse sequence duration $\tau_{pp}$ and phase $\phi$, plotted for five increasing exchange rates ($k_d$). As $k_d$ increases from 1.0 s$^{-1}$ to 300 s$^{-1}$, the resonance condition (visible as high-intensity red/blue bands) broadens and shifts.
The black dashed lines represent the theoretical resonance condition for the $n=0$ harmonic, derived from the coherent evolution formula $\tau_{pp} = \pm \phi / (\pi J)$. The close alignment between the simulated high-polarisation regions and these analytic lines confirms that the coherent transfer mechanism dominates, particularly at lower exchange rates. At higher $k_d$, the resonance features become more diffuse, indicating the interplay between coherent evolution and exchange dynamics.}
  \label{fgr:SABRE_PP_Phase_Tau_Heatmaps} 
\end{figure}

Nevertheless, significant polarisation gains are achievable in the high-$k_d$ regime by deviating from the standard coherent condition, as seen in Figure~\ref{fgr:PP_coherent_vs_optim}(b, c). To maximize transfer in this regime, the phase shift must be increased beyond the theoretical prediction, as illustrated in Figure~\ref{fgr:PP_coherent_vs_optim}(d). This deviation suggests that the resonance condition can be further optimized locally—for instance, by iteratively tuning $\phi$ and $\tau$ around the theoretical maximum—to better account for exchange dynamics (see Figure~\ref{fgr:SABRE_PP_Phase_Tau_Heatmaps}).

Our simulations predict an optimal pulse duration $\tau_{\text{PP}}$ in the range of 15–25 ms, which is consistent with experimentally observed resonance conditions, see Table~\ref{table:SABRE-PulsePol}. Furthermore, the model indicates that the optimal phase depends strongly on the exchange dynamics, shifting from approximately -36$^{\circ}$ for slowly exchanging species (e.g., acetonitrile and pyridine) to -72$^{\circ}$ for rapidly exchanging substrates (e.g., metronidazole).

\subsection{Overall comparison and coherent dynamics}

\begin{figure}[H]
\centering
  \includegraphics[width=0.9\textwidth]{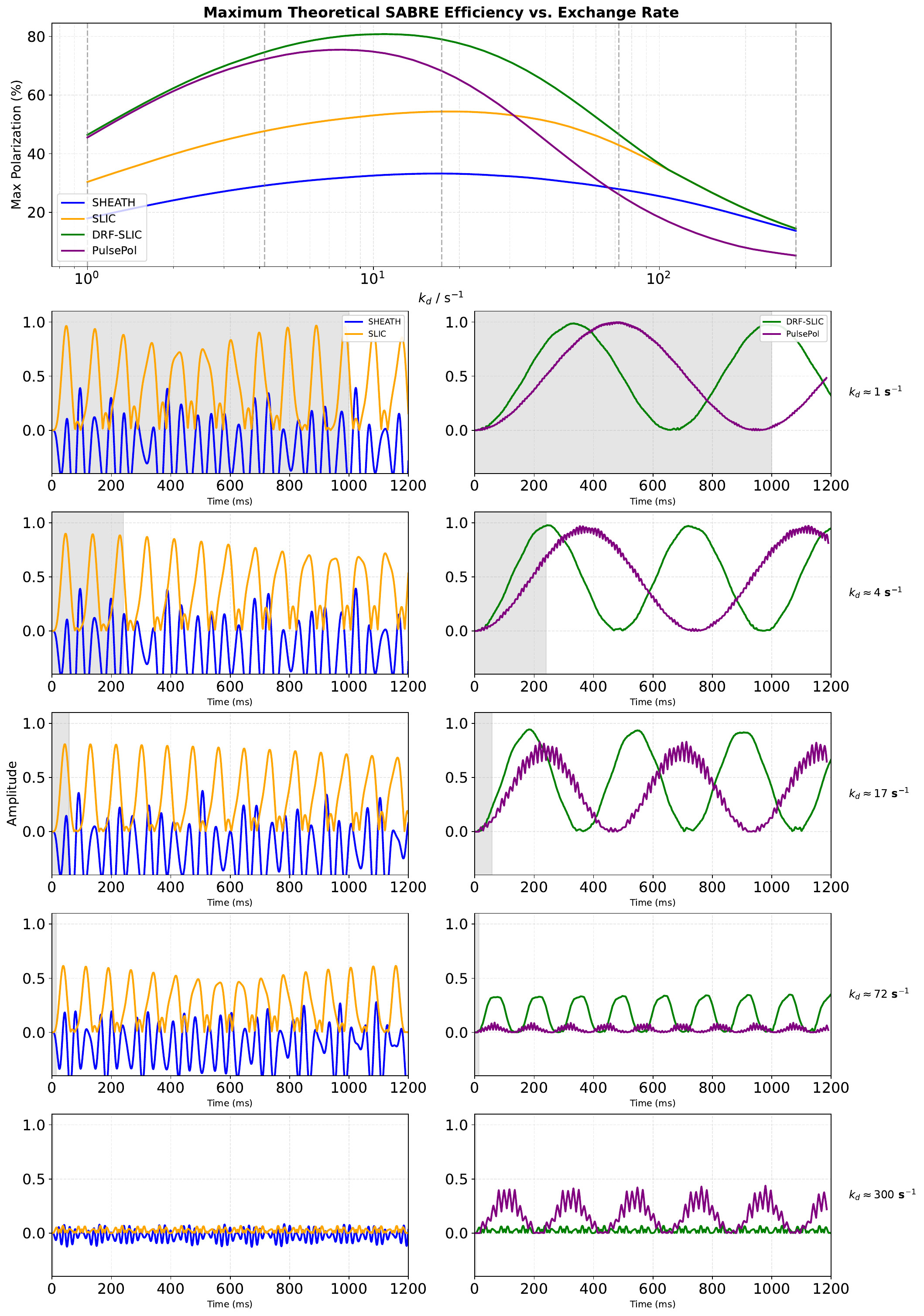}
  \caption{Theoretical limits and coherent evolution of SABRE methods.
(Top) Maximum $^{15}$N polarisation vs. ligand exchange rate ($k_d$) for SABRE-SHEATH (blue), SLIC (orange), DRF-SLIC (green), and PulsePol (purple). Vertical dashed lines indicate the five specific $k_d$ values analyzed in the panels below.
(Bottom) Time-dependent polarisation buildup for the selected exchange rates (rows). The left column compares SHEATH and SLIC methods, while the right column compares DRF-SLIC and PulsePol methods, proposed in this article. The grey shaded region represents the complex lifetime ($t \le 1/k_d$). All simulations employ optimized control parameters.}
  \label{fgr:all_in_all} 
\end{figure}

Figure~\ref{fgr:all_in_all} presents an overall comparison of the maximum achievable polarisation across different methods. Supplementing the main text, the coherent evolution using optimal SABRE parameters is shown below for various $k_d$ values. It is evident that DRF-SLIC and PulsePol outperform SLIC and SHEATH primarily due to their ability to adjust the evolution frequency—which decreases with lower $k_d$ values—while maintaining maximal transfer amplitude.

\bibliography{refs} 
\bibliographystyle{rsc} 